\documentclass[12pt,english]{paper}
\usepackage[T1]{fontenc}
\usepackage[latin9]{inputenc}
\usepackage{babel}
 
\usepackage{graphicx}
 
\usepackage{amsmath}
\usepackage{amsfonts}
\usepackage{amsthm}
\usepackage{mathtools}
\usepackage{amssymb}
\usepackage{bbm}
 
\usepackage{color} 
\usepackage{hyperref}
 
\usepackage[normalem]{ulem}
\usepackage{natbib}

 \usepackage{cancel}
\usepackage{geometry}
\usepackage{subcaption}

\usepackage{nicematrix}
\usepackage{booktabs}

\usepackage{tabularx}

\usepackage[colorinlistoftodos]{todonotes} 

\makeatletter

\theoremstyle{plain}

\theoremstyle{plain}
\newtheorem{lemma}{\protect\lemmaname}
\theoremstyle{plain}
\newtheorem{proposition}{\protect\propositionname}
\theoremstyle{definition}
\newtheorem{defn}{\protect\definitionname}
\theoremstyle{plain}
\newtheorem{ass}{\protect\assumptionname}
\theoremstyle{plain}

\theoremstyle{plain}
\newtheorem{remark}{\protect\remarkname}

\usepackage{tikz}
\usetikzlibrary{patterns}

\makeatother

\providecommand{\definitionname}{Definition}
\providecommand{\lemmaname}{Lemma}
\providecommand{\propositionname}{Proposition}
\providecommand{\theoremname}{Theorem}
\providecommand{\assumptionname}{Assumption}
\providecommand{\corollaryname}{Corollary}
\providecommand{\remarkname}{Remark}

\def\ubeta{\underline\beta}
\def\udelta{\underline\delta}

\newif\ifanonymous
\anonymousfalse  

\begin{document}

\title{\vspace{-3.5cm} Fair Combinatorial Auctions: Endogenous Best Execution in Blockchain Trade-Intent Markets%
\ifanonymous\else
\thanks{This paper initially circulated with the title ``Combinatorial Auctions without a Numeraire: The Case of Blockchain Trade-Intent Auctions.'' We are grateful to Felix Leupold and Haris Angelidakis for the initial discussions that led to the writing of this paper. We also thank Patrick Bolton, Eric Budish, Tim Roughgarden, and participants at the Paris Dauphine ``Tech for Finance'' 2024 conference, the 2024 Lisbon Meeting in Game Theory and Applications,  ICCF24, Devcon 7 SEA, the 2024 QRFE Workshop on Blockchain-based markets Fintech and Cryptocurrencies, and Columbia CryptoEconomics Workshop 2024, CBER Webinar,  for numerous comments and suggestions. }%
\fi
}


\ifanonymous
\author{}
\else
\author{Andrea Canidio\thanks{Category Labs, email: acanidio@category.xyz (corresponding author)} and Felix Henneke\thanks{CoW Protocol.}}
\fi

\maketitle

\noindent This version: \today 

\begin{abstract}

Trade-intent auctions intermediate around USD~9~billion in monthly trading volume. In these auctions, specialized intermediaries called solvers compete for the right to execute  orders across fragmented blockchain-based financial markets. These auctions are combinatorial because executing multiple trade intents jointly generates additional efficiencies. However, there is no best-execution benchmark to determine how to share those efficiencies: the best possible execution of a trade is solvers' private information and must be elicited. We study theoretically the two main mechanisms: batch auctions, in which a group of trades is auctioned off jointly, and independent trade-by-trade auctions. Batch auctions return more total value to traders, but their outcome may be unfair, in the sense of leaving one trader worse off than under independent auctions. We propose a \emph{fair combinatorial auction}: solvers bid on individual trades and on batches of trades, but a batched bid is filtered out if any trader earns less than an execution benchmark constructed from the bids on individual trades and a counterfactual mechanism. Whether fairness guarantees arise in equilibrium depends on the counterfactual mechanism: independent first-price auctions generate such guarantees; independent second-price auctions do not. These fairness guarantees come at a cost: a lower total value returned to traders. 


\end{abstract}

\section{Introduction}

Modern financial markets increasingly rely on intermediaries to execute complex trades across fragmented venues. This fragmentation is particularly pronounced on permissionless blockchains such as Ethereum, where anyone can create a new financial market at negligible cost. Dozens of such markets coexist, differing in their mechanisms, fees, and liquidity. Executing a trade efficiently may therefore require accessing several markets, often through intermediate assets. 


Trade-intent auctions emerged in response to these execution challenges.\footnote{A further motivation for trade-intent auctions is protection from ``sandwich attacks,'' in which an adversary front-runs and back-runs a trade, worsening its execution terms. In trade-intent auctions, the execution of a trade is delegated to specialized entities, who employ sophisticated strategies to prevent such attacks.} A \emph{trade intent} is a partially specified blockchain transaction: it defines the asset to sell and the asset to buy, while leaving the execution route unspecified. An auction allocates the right to complete this transaction. In it, specialized entities called \emph{solvers} compete by proposing the most favorable execution for each intent. The earliest implementation is the \emph{batch auction} used by CoW Protocol (whose primary interface is CoW Swap): a group of trade intents is auctioned simultaneously, each solver submits a single batched bid covering multiple intents, and the auctioneer selects one winning solver. Subsequent entrants, such as 1inch (Fusion) and Uniswap (UniswapX), instead run a separate descending-price (Dutch) auction for each trade intent. 
Between June 2025 and May 2026, on Ethereum alone, they intermediated an average of USD~8.7~billion in monthly volume, peaking at USD~14.3~billion in August 2025.\footnote{Source: \url{https://dune.com/queries/3058985/5091785}. Following completion of the first draft of this paper, CoW Protocol announced that it would replace its batch auction with the fair combinatorial auction introduced later in this paper.} By some accounts, when bot-generated transactions are excluded, the majority of Ethereum trading volume is routed via trade-intent auctions.\footnote{Flashbots maintains a dashboard of Ethereum trading volume submitted through known user-facing frontends (i.e., websites or apps), therefore excluding bot-generated transactions. Trade-intent auction frontends (CoW Protocol, 1inch, and UniswapX) account for more than half of this volume. See \url{https://orderflow.art/frontends}.}

Trade-intent auctions share features with several mechanisms from traditional finance. The relation between a trade intent and a blockchain transaction is analogous to that between a parent order and its child orders: a parent order specifies the client's intention (e.g., to sell a stock within a given period), which a broker then decomposes into child orders routed to specific markets. Parent orders, however, are typically not auctioned. The closest analogue is therefore the SEC's proposed Rule 615 (the ``order competition'' rule), which would have required platforms such as Robinhood to select the executing broker for each retail order through an auction, rather than routing all orders to the same intermediary (see also \citealp{ernst2023would}). Trade-intent auctions also resemble foreign-exchange markets: users can exchange any asset for any other asset, so there is no numeraire.\footnote{There is, however, an important difference in scale: roughly 100 currencies trade in foreign-exchange markets, and the top five pairs --- all involving the USD --- account for around 65\% of total volume; each month, CoW Protocol on Ethereum alone facilitates trades among roughly 2,000 distinct assets, and the top five pairs account for about 35\% of total volume.}  

To compete in a trade-intent auction, a solver must specialize in particular trade intents while also exploiting complementarities across different trade intents. Solvers specialize in different ways: some develop sophisticated algorithms to access public marketplaces, others rely on their own inventory, and still others focus on assets requiring specialized operations.\footnote{Blockchain-based 
assets such as cryptocurrencies and tokens are ``programmable money'' whose properties can vary widely.} To exploit complementarities, solvers may identify coincidences of wants that allow traders to transact directly, sometimes through complex ring or multi-step trades. Moreover, because blockchain fees (``gas costs'') are largely independent of trade volume, executing multiple trades jointly reduces total fees. 

It follows that trade-intent auctions are inherently combinatorial: the optimal execution of a trade depends on the other trades executed by the same solver. As in most combinatorial auctions, the designer must balance the benefits of specialization (allocating each trade intent to the best solver) against the additional efficiencies from batching (assigning multiple orders to the same solver). Unlike standard combinatorial auctions, however, a new concern arises: \textit{fairness}. Traders may demand different assets, the efficiencies from batching may materialize in any asset, and there is no best-execution benchmark to discipline how those efficiencies are shared.

As an example, consider two traders, one requesting \textit{DOGE} (a cryptocurrency with a market capitalization of about USD 23 billion as of mid-2026) and the other requesting \textit{ETH} (the native currency of Ethereum, with a market capitalization of about USD 380 billion). Two solvers each submit a bid for the batch. Suppose the first solver's bid yields more \textit{DOGE}, fewer \textit{ETH}, and a higher overall market value than the second solver's bid. Because only one trader demands \textit{DOGE}, the auctioneer cannot observe how much (if at all) the other trader values this asset, and thus cannot transfer utility between traders by reallocating the extra \textit{DOGE}. The auctioneer or solvers could instead trade \textit{DOGE} for \textit{ETH}, yet frictions and fees may still render the second solver's bid preferable for the trader requesting \textit{ETH}. The traders may therefore disagree on which bid should win.  

This paper develops a theoretical model of trade-intent auctions. Two traders each demand a different asset, and two solvers can execute them. The model features both gains from specialization --- each solver has an absolute advantage in executing one trade --- and complementarities --- a solver generates more of an asset when it also executes the other trade. We assume notional market prices exist for the demanded assets, so bids can be valued in dollars. However, the assets are illiquid \textit{within the auction}: exchanging them for one another or for dollars involves frictions and fees that are solver-specific and privately known --- a restriction we formalize as a \textit{feasibility constraint} on the amounts each solver can return.  As in the above example, the main issue facing the designer is how to share the additional efficiency from executing multiple trades together as a batch.  The designer thus faces a two-dimensional elicitation problem: it must elicit from solvers not only the joint best execution of the trades, but also the best-execution benchmark against which the fairness of the joint execution is judged.

We use the model to study the two dominant auction formats: batch auctions and independent standard auctions. Intuitively, independent auctions better exploit specialization, since different solvers may win different trades, whereas batch auctions better exploit complementarities across trades. This intuition, however, overlooks an important point. When specialization is strong, competition in independent auctions is weak: the stronger solver in each auction can win while returning little value to traders. Batching instead forces all solvers to compete directly, strengthening bidding incentives and increasing the amounts returned to traders. Batch auctions, however, have a central drawback. Solvers' payoffs depend only on the total value returned, not on its distribution. Multiple equilibria therefore arise, and the gains from batching may accrue disproportionately to one trader --- an outcome practitioners recognize as \emph{unfair}.\footnote{CoW Protocol attempts to mitigate this problem with a constraint analogous to a best-execution requirement: it compares each solution to a benchmark constructed using an in-house routing algorithm and requires solvers to reimburse users when execution falls short. CoW Protocol refers to this benchmark as EBBO (Ethereum Best Bid and Offer; see \texttt{https://docs.cow.fi/cow-protocol/reference/core/auctions/ebbo-rules}). This benchmark, however, does not incorporate solvers' private information, and therefore fails to reflect the true best execution of an individual order. } 

Motivated by this observation, we formalize a notion of fairness and introduce a new auction format: the \textit{fair combinatorial auction}, whose goal is to elicit a best-execution benchmark directly from solvers. In this auction, solvers bid both on individual trades and on batches of trades. Our notion of fairness draws on cooperative game theory: an outcome is fair if and only if every trader in a batch is at least as well off as under a reference outcome. In cooperative games, this reference is typically the players' ``outside options'' --- the highest payoff each can achieve without cooperation. In our setting, the reference is instead the outcome of the auction when only bids on individual orders are considered.\footnote{If multiple batched bids dominate the reference outcome for all traders, the notion of fairness can be strengthened, for example, by ranking allocations by the product of traders' gains relative to the benchmark. Our main results hold under any such strengthening.} The fairness benchmark is therefore constructed endogenously from solvers' bids. To compare the fair combinatorial auction with existing mechanisms, we assume it can be decomposed into two individual-trade auctions --- either first or second price --- and a batched auction in which the winner ``pays its bid,'' returning to traders the amounts specified in its batched offer. Fairness then requires the auctioneer to ignore any batched bid that delivers less than the outcome constructed from the individual-trade bids.

The fair combinatorial auction introduces two novel strategic considerations. First, a solver's individual-trade bids influence which batched bids are deemed fair. Solvers may therefore bid more aggressively on individual trades than in independent standard auctions, hoping to render the opponent's batched bid unfair and have it excluded. Second, anticipating this exclusion, a solver expects weaker competition for the batch as a whole and may bid less aggressively in the batched auction. The strength of each effect depends on the format of the individual-trade auctions: first price or second price.

If the individual-trade auctions are in second price, each solver has limited ability to influence the fairness benchmark. A solver bidding above the opponent cannot raise the benchmark; a solver bidding below can raise it only up to the opponent's bids. In particular, if both solvers set their individual-trade bids to zero --- effectively abstaining from individual bidding --- the outcome is always batching. The equilibrium therefore delivers no fairness guarantee: the two strategic effects described above are moot, and the set of equilibrium outcomes is identical to that of the batch auction.

If instead the individual-trade auctions are in first price, each solver can influence the fairness benchmark through its bids. Both strategic effects are now active: solvers bid more aggressively in the individual-trade auctions but less aggressively in the batched auction. The equilibrium is unique (up to payoff-irrelevant bids), and it falls into one of three regimes:
\begin{itemize}
    \item \textbf{Specialization.} When differences in solvers' productivity across trades are large, each solver disqualifies the opponent's batched bid as unfair. Each trader then receives the assets that the less efficient solver at its order would provide \textit{when matched with both orders}. This qualifier matters, because it implies that, even under specialization, both traders receive strictly more than in the equilibrium of the independent standard auctions. 
    \item \textbf{Competitive batching.} When the benefit of batching is large relative to solvers' productivity differences, the equilibrium resembles that of the batch auction, with one important difference: each trader receives \textit{at least} the assets that the less efficient solver at its order would provide when matched with both orders.  
    \item \textbf{Uncompetitive batching.} When the benefit of batching is in an intermediate range and one solver is much more productive than the other, that solver can easily disqualify the opponent's batched bid as unfair. It then wins both orders with a relatively low bid. Again, each trader receives the assets that the less efficient solver at its order would provide when matched with both orders.
\end{itemize}
The auction therefore provides strong fairness guarantees: in all regimes, each trader receives strictly more than in the equilibrium of the independent first-price auctions. However, these guarantees come at a cost: the total market value of assets delivered to traders is weakly lower than in a batch auction --- strictly so under specialization and uncompetitive batching.

Comparing the two versions of the fair combinatorial auction --- individual-trade auctions in first versus second price --- yields two main insights. First, the payoff each trader is guaranteed in equilibrium can differ sharply from the payoff in the equilibrium of the benchmark mechanism itself. When the benchmark is two second-price auctions, a trader may receive \textit{less} in the fair combinatorial auction than in the equilibrium of those auctions. When the benchmark is two first-price auctions, every trader receives more than in the equilibrium of the benchmark mechanism. The endogenously constructed reference for fairness therefore does not coincide with a traditional best-execution guarantee: depending on the benchmark, it may be more or less favorable than the outcome of a single-trade auction. Second, the comparison reveals a variant of the classic fairness--efficiency trade-off: stronger fairness guarantees come at the cost of lower aggregate market value returned to traders. This is the price of constructing the fairness reference endogenously within the mechanism. In an extension, we embed the mechanism in a one-parameter family spanning the range between the batch auction and the fair combinatorial auction, and show that the latter is the unique member of the family whose fairness guarantees do not depend on details of the environment that the designer does not know.

The remainder of the paper is organized as follows. Next, we discuss the 
relevant literature. Section~\ref{sec: model} introduces the model, which we then use in 
Section~\ref{sec: standard_auctions} to analyze the two dominant auction formats. Section~\ref{sec: fairness} introduces our notion of fairness, and 
Section~\ref{sec: equilibrium} derives the equilibrium of the fair combinatorial auction. 
Section~\ref{sec: discussions} discusses extensions, and the final section concludes. 
All mathematical derivations omitted from the text appear in the Appendix.

\subsection{Relevant literature}

The problem we study connects to classic market-design questions in finance, including the use of auctions for price discovery and market structuring (e.g., \citealp{madhavan2000price}; \citealp{10.1093/qje/qjv027}). This literature typically analyzes auctions within a single market: \cite{10.1093/qje/qjv027}, for example, propose frequent batch auctions to eliminate latency arbitrage on a single exchange, whereas our batch auctions bundle heterogeneous trades --- orders that may have no asset in common --- to exploit execution complementarities. 

Trade-intent auctions operate \emph{across} markets: their goal is to aggregate liquidity from multiple venues. In this sense, our work is conceptually closer to the literature on order-flow internalization and routing (e.g., \citealp{easley1996cream}). Within this strand, \cite{ernst2023would} compare the current system, in which brokers bid on a stream of orders, with the SEC's proposed Rule 615, which would hold a separate auction for each order. They find that bidding on a stream of orders increases competition among bidders at the cost of ex post allocative inefficiency --- a trade-off analogous to the one we identify when comparing batch auctions with independent standard auctions. Because their environment is sequential rather than combinatorial, however, they do not confront fairness considerations or explore combinatorial designs.

From a theoretical standpoint, the problem we study resembles multi-unit or combinatorial assignment problems because there is no numeraire (see \citealp{budish2011combinatorial}, \citealp{budish2012multi}). However, assignment problems exclude monetary transfers. In trade-intent auctions, by contrast, multiple \textit{types} of monetary transfers exist, and some traders may accept certain transfers but not others. The assignment literature has also proposed several definitions of fairness, capturing whether an agent ``envies'' another agent's equilibrium allocation. Fairness is thus defined by comparing agents \textit{within the same mechanism}. Our notion of fairness instead compares the outcome of a mechanism with that of a \textit{counterfactual mechanism}.

Our definition of fairness draws on cooperative game theory. In our setting, players' outside options correspond to separate trade-by-trade auctions, and ``cooperation'' corresponds to batching trades together. Cooperation then arises only if each player earns at least as much from cooperating as from its outside option. This 
literature distinguishes between games that allow frictionless utility transfers across players (transferable-utility or TU games) and those that do not (non-transferable-utility or NTU games). The problem we study is a matching problem with non-transferable utility, related to \citet{legros2007beauty}, who consider environments where transfers are possible but subject to frictions. The key difference is informational. In \citet{legros2007beauty}, and in the cooperative-game literature more generally, both the production function and the fairness benchmark (the outside options) are common knowledge. In our setting, both the solvers' technologies and the fairness benchmark must be elicited as part of the mechanism itself.

Within the auction literature, the most closely related paper is \citet{che1998standard}, who study a single auction with financially constrained bidders and show that a first-price auction generates higher revenue than a 
second-price auction. The intuition is that bids are lower in a first-price auction, so the constraint is less likely to bind. Their insight relates to ours: in our model, market frictions also impose a constraint on feasible bids, which we formalize as a feasibility constraint. The difference is that our constraint depends on the assignment of trade intents to solvers and is therefore relevant only in a combinatorial environment.\footnote{\citet{dobzinski2012multi} study multiple simultaneous auctions with financially constrained bidders, but in their model a single budget constraint applies to the sum of all bids. Our feasibility constraint instead binds asset by asset and depends on the allocation of trades across solvers.} Our paper also relates to the literature on multiple simultaneous auctions, including 
\citet{milgrom2000putting}, \citet{krishna1996simultaneous}, \citet{goeree2014equilibrium}, and \citet{GENTRY2019109}. In particular, \citet{GENTRY2019109} analyze simultaneous sealed-bid auctions with complementarities, one of the benchmarks we study. Like theirs, our equilibrium constructions rely on the endogenous tie-breaking approach of \citet{jackson2002communication}, which plays a role at several points of our analysis.

We also contribute to the literature on decentralized finance (DeFi). Existing research has focused primarily on individual on-chain markets, most notably automated market makers (AMMs) (e.g., \citealp{lehar2025decentralized, capponi2025liquidity, milionis2022automated, hasbrouck2025economic}). Trade-intent auctions have received far less attention. \cite{chitra2024analysis} consider a single Dutch auction and study how the equilibrium changes with the number of bidders. We instead fix the number of bidders and analyze how different mechanisms perform when trades exhibit complementarities and the designer faces fairness considerations. \cite{bachu2024quantifying} and \cite{yuminaga2025executionwelfaresolverbaseddexes} empirically study the implementation of trade-intent auctions in 1inch Fusion, UniswapX, and CoW Protocol, but do not analyze their mechanism-design properties.

Finally, this paper contributes to the emerging literature at the intersection of blockchain 
and mechanism design (see \citealp{townsend2020distributed} for an overview). Most of this
literature examines how blockchain technology generates commitment and thereby enables the 
implementation of mechanisms (see \citealp{holden_malani_2021}; \citealp{gans2019fine}; 
\citealp{lee2021optimal}; \citealp{brzustowski2021smart}; \citealp{townsend-zhang}; 
\citealp{canidio2022Auction-pp}; \citealp{canidio2023auctions}). A smaller but 
more closely related strand applies mechanism-design insights to improve blockchain 
protocols (see, for example, \citealp{gans2022mechanism}; \citealp{gans2023solomonic}; 
\citealp{Gans2023}; \citealp{canidio-danos}). Our paper adds 
to this second strand by proposing a new design for blockchain-based trade-intent auctions.

\section{The model}\label{sec: model}

There are two traders. Trader~1 wishes to sell one unit of asset~$A$ for asset~$B$, and Trader~2 wishes to sell one unit of asset~$C$ for asset~$D$. Two solvers can execute these trades by exchanging the assets in external financial markets or by using their own inventories. All assets have notional market prices, but exchanging them is subject to frictions and fees that are solver-specific, depend on the set of trades the solver executes, and are privately known. These frictions define each solver's \textit{execution frontier}, which we introduce formally below. 

An auction allocates trades to solvers. In it, solvers bid by proposing prices to the traders. A solver's profit is the difference between the prices its execution frontier allows and the prices it proposes to the traders. This difference is a form of spread. Unlike a quoted spread, however, it is determined endogenously within the auction and depends on the specific auction mechanism.

The key departure from standard auction models is that assets~$B$ and~$D$ cannot be exchanged within the auction, which we formalize in Assumption~\ref{ass: feasibility}. If $B$ and $D$ could be traded frictionlessly, both traders would effectively demand the same asset, and the multi-asset, combinatorial structure of the problem would disappear (see Section~\ref{sec: benchmark} for discussion).

\paragraph{Execution frontier.}
To parsimoniously capture both the benefit of specialization and the additional efficiencies from batching, 
we specify the following execution frontier.  Let $\beta$ and $\delta$ be independently drawn 
at the beginning of the game from continuous and atomless distributions on 
$[\underline{\beta}, \overline{\beta}]$ and $[\underline{\delta}, \overline{\delta}]$, respectively, 
with $0 < \underline{\beta} < \overline{\beta}$ and $0 < \underline{\delta} < \overline{\delta}$. 

\begin{itemize}
\item If solver~1 executes only order~1, it can deliver up to $\beta$ units of asset~$B$. 
If solver~2 executes only order~1, it can deliver up to $\underline{\beta}$ units of asset~$B$.

\item If solver~2 executes only order~2, it can deliver up to $\delta$ units of asset~$D$.  
If solver~1 executes only order~2, it can deliver up to $\underline{\delta}$ units of asset~$D$.

\item If solver~1 executes both orders, it can deliver up to $g \cdot \beta$ units of asset~$B$ and 
$g \cdot \underline{\delta}$ units of asset~$D$; if solver~2 executes both orders, it can deliver up to 
$g \cdot \underline{\beta}$ units of asset~$B$ and $g \cdot \delta$ units of asset~$D$.
\end{itemize}
These quantities are reduced-form execution frontiers summarizing routing ability, access to liquidity, inventory, gas optimization, the ability to internalize offsetting trades, and are net of execution costs. Only the winning solver receives the right to execute a trade intent, so a solver's frontier matters only for trades it wins.

Thus, each solver is specialized in one order: solver~1 in order~1 and solver~2 in order~2. At the same time, there are additional efficiencies when the same solver executes both trades, measured by the parameter $g \ge 1$. For example, exchanging $A$ for $B$ and $C$ for $D$ may require routing through one or more intermediate assets. When these routes overlap, a solver executing both trades can share intermediate steps, saving on gas costs, or bypass some fees altogether through direct peer-to-peer matching along those steps. The trade-off between specialization and the additional efficiencies from batching is present if $g \ubeta<\overline{\beta}$ and  $g \udelta<\overline{\delta}$, which we therefore assume. 

Finally, we abstract from trade-offs in routing: when executing both orders, a solver 
cannot generate more of one asset at the expense of the other (as we will see, a solver \textit{can} return more than one asset at the expense of the other when bidding). We discuss the consequences of relaxing this assumption in Section~\ref{sec: technologically unfair}. 

\paragraph{Feasibility constraint.}
Let a matching of trades and solvers be denoted by 
\[
\mu  \in \left\lbrace (1,2), (2,1), (1,1), (2,2) \right\rbrace
\]
where $\mu=(i,j)$ indicates that solver $i\in\{1,2\}$ is matched with order~$1$ and solver $j\in\{1,2\}$ with order~2. 
Let $x_i(B)$ and $x_i(D)$ denote the amounts of assets $B$ and $D$ returned by solver~$i\in\{1,2\}$. 
Our main assumption is that solvers face two asset-specific feasibility constraints (instead of a single budget constraint as in \citealp{che1998standard}).

\begin{ass}[Feasibility constraint]\label{ass: feasibility}
Let $\eta(B) \ge 0$ and $\eta(D) \ge 0$ denote the inventories of assets $B$ and $D$ 
held by solver~1 and solver~2, respectively. These inventories are privately known to each solver. 
The following constraints must hold:
\begin{itemize}
    \item If $\mu = (1,1)$, then $x_1(B) \le g\beta + \eta(B)$, 
    $x_1(D) \le g\underline{\delta}$, $x_2(B) = 0$, and $x_2(D) \le \eta(D)$.
    \item If $\mu = (2,2)$, then $x_2(B) \le g\underline{\beta}$, 
    $x_2(D) \le g\delta + \eta(D)$, $x_1(B) \le \eta(B)$, and $x_1(D) = 0$.
    \item If $\mu = (1,2)$, then $x_1(B) \le \beta + \eta(B)$, 
    $x_2(D) \le \delta + \eta(D)$, and $x_1(D) = x_2(B) = 0$.
    \item If $\mu = (2,1)$, then $x_2(B) \le \underline{\beta}$, 
    $x_1(D) \le \underline{\delta}$, $x_1(B) \le \eta(B)$, and $x_2(D) \le \eta(D)$.
\end{itemize}
\end{ass}

The main implication of the assumption is that, within the auction, solvers cannot convert asset $B$ into $D$ (or vice versa) after trading $A$ for $B$ and $C$ for $D$. It also implies that each solver may use its inventory only to fill the order in which it has an advantage, but no solver holds both assets and thus cannot act as a market maker in the $B$-$D$ pair. Solvers therefore cannot reallocate utility between traders. If they could, the problem would collapse to a standard one-dimensional auction (see Section~\ref{sec: benchmark}). 

Finally, note that the auction outcome is stochastic because each solver is uncertain about the other solver's action. The feasibility constraint must hold in every outcome that occurs with strictly positive probability.

\paragraph{Information.}
The parameters $\underline{\beta}$, $\overline{\beta}$, $\underline{\delta}$, $\overline{\delta}$, and $g$ are common knowledge among the players but unknown to the designer, who cannot condition the mechanism on their realizations. By contrast, $\beta$ is privately known to solver~1 and $\delta$ is privately known to solver~2.

\paragraph{Payoffs.}
Let $p_{DB}$ denote the notional market price of asset~$D$ in terms of asset~$B$, assumed common 
knowledge. For $(x_1(B),x_1(D))$ satisfying Assumption~\ref{ass: feasibility}, solver~1's payoff 
is
\[
\pi_1(\mu)=
\begin{cases}
g\beta - x_1(B) + p_{DB}\,(g\underline{\delta} - x_1(D)) & \text{if } \mu = (1,1),\\[3pt]
\beta - x_1(B) & \text{if } \mu = (1,2),\\[3pt]
-\,x_1(B) + p_{DB}\,(\underline{\delta} - x_1(D)) & \text{if } \mu = (2,1),\\[3pt]
-\,x_1(B) & \text{if } \mu = (2,2).
\end{cases}
\]
Note that for $\mu = (2,1)$ and $\mu = (2,2)$, solver~1 does not execute order~1: any delivery $x_1(B)$ must come from its inventory, at cost $x_1(B)$. Similarly, for $(x_2(B),x_2(D))$ satisfying Assumption~\ref{ass: feasibility}, solver~2's payoff is
\[
\pi_2(\mu)=
\begin{cases}
-\,p_{DB}\,x_2(D) & \text{if } \mu = (1,1),\\[3pt]
p_{DB}\,(\delta - x_2(D)) & \text{if } \mu = (1,2),\\[3pt]
\underline{\beta} - x_2(B) - p_{DB}\,x_2(D) & \text{if } \mu = (2,1),\\[3pt]
g\underline{\beta} - x_2(B) - p_{DB}\,(g\delta - x_2(D)) & \text{if } \mu = (2,2).
\end{cases}
\]
Each solver values asset~$D$ at the notional market price $p_{DB}$, even though exchanging $D$ 
for $B$ within the auction is infeasible. This assumption reflects the timing of trade: solvers 
must deliver assets to traders immediately, but they can later rebalance their inventories in 
external markets over a longer horizon, reducing (here, for simplicity, eliminating) frictions and fees.\footnote{Some protocols, such as CoW Protocol and 1inch, also distribute rewards to solvers to attract participation in the auction, so a solver's payoff depends on both the tokens retained and the rewards received. Because different protocols use different reward formulas, we abstract away from rewards and focus on the key auction dynamics: how competition among solvers elicits private execution opportunities and determines how much of the resulting surplus is passed through to traders.}

Traders' payoffs are simply the assets they receive: if Trader~1 receives $x$ units of asset~$B$ 
and Trader~2 receives $y$ units of asset~$D$, their utilities are $u_1 = x$ and $u_2 = y$, 
respectively.

\paragraph{Timing.}
The timing of the game is as follows:
\begin{enumerate}
    \item The productivity parameters $\beta$ and $\delta$ are drawn.
    \item Each solver $i \in \{1,2\}$ simultaneously submits three bids. The first two, 
    $q_i(B)$ and $q_i(D)$, are bids on the individual orders, denominated in units of assets 
    $B$ and $D$, respectively. The third is a tuple 
    $Q_i = \{Q_i(B), Q_i(D)\}$, representing a bid for the entire batch of both orders, 
    where $Q_i(B)$ and $Q_i(D)$ are in units of assets $B$ and $D$.
    \item The mechanism determines the matching $\mu$ and specifies the quantities of each 
    asset to be delivered to the traders.
\end{enumerate}

\paragraph{Mechanism.}
A message profile is defined as
\[
m \equiv \{q_1(B), q_1(D), Q_1(B), Q_1(D), q_2(B), q_2(D), Q_2(B), Q_2(D)\} \in \mathbb{R}_+^8.
\]
A mechanism is then $\Gamma = \{f, b_1, b_2, d_1, d_2\}$, where
\begin{itemize}
    \item $f: \mathbb{R}_+^8 \to \{1,2\} \times \{1,2\}$ determines the matching of trades to solvers;
    \item $b_i: \mathbb{R}_+^8 \to \mathbb{R}_+$ specifies the quantity of asset~$B$ delivered by solver~$i$ to Trader~1;
    \item $d_i: \mathbb{R}_+^8 \to \mathbb{R}_+$ specifies the quantity of asset~$D$ delivered by solver~$i$ to Trader~2.
\end{itemize}
The mechanism is common knowledge at the beginning of the game.

An important observation is that the feasibility constraint limits the set of admissible 
mechanisms. Any \emph{all-pay} auction is infeasible, as the solver losing a trade could still 
be required to deliver assets to the corresponding trader. Moreover, Vickrey-Clarke-Groves (VCG) 
mechanisms are also incompatible with feasibility. In a VCG mechanism, each winner pays the 
highest-valued combination of bids excluding her own. If the equilibrium is specialization, removing one solver's bid would lead to batching, so that a solver winning only one order would be required to deliver 
both assets---a direct violation of feasibility. The following definition formalizes this intuition.

\begin{defn}[Feasible mechanism]
A mechanism $\Gamma$ is \emph{feasible} if, for each solver $i \in \{1,2\}$, there exist bids 
$q_i(B) > \underline{\beta}$, $q_i(D) > \underline{\delta}$, $Q_i(B) > \underline{\beta}$, and $Q_i(D) > \underline{\delta}$ such that, for any bids 
$\{q_{-i}(B), q_{-i}(D), Q_{-i}(B), Q_{-i}(D)\}$ submitted by the other solver 
$-i \in \{1,2\} \setminus \{i\}$, the resulting allocation $f(m)$ and deliveries 
$b_i(m)$ and $d_i(m)$ satisfy Assumption~\ref{ass: feasibility} for $\eta(B) = \eta(D) = 0$.
\end{defn}

In any feasible mechanism, therefore, when the outcome is specialization each winning solver delivers assets 
to only one trader: for any message profile $m \in \mathbb{R}_+^8$ such that 
$f(m) = (i, -i)$, we must have $b_{-i}(m) = d_i(m) = 0$. Similarly, if a solver wins no trades 
($i \notin f(m)$), it delivers no assets, so that $b_i(m) = d_i(m) = 0$.

\paragraph{Order Cancellations.} To guarantee the existence of the equilibrium of the independent standard auctions (see the discussion in Section \ref{sec: independent standard}), we perturb the game by assuming that there is an arbitrarily small probability that a solver may fail to win the order on which it has a comparative advantage. We do so by allowing traders to cancel their order: with probability $\varepsilon > 0$, an order may be canceled after the auction but before execution --- in practice, a trader may withdraw the tokens to be sold from its wallet after submitting the trade intent. Bids are contingent commitments: an individual-trade bid is void if the corresponding order is canceled, and a batched bid is void if any order in the batch is canceled, in which case the remaining order is resubmitted to a subsequent auction. The payoff introduced earlier and all results below are stated conditional on no order being canceled, and none depends on the value of $\varepsilon$:  the perturbation plays no role in our results beyond guaranteeing the existence of the equilibrium of the independent standard auctions.

\section{Examples: Batch auctions and independent standard auctions}\label{sec: standard_auctions}

We now use the framework to analyze the two dominant mechanisms for trade-intent auctions: independent standard auctions and batch auctions. Beyond their empirical relevance (see the Introduction), these mechanisms also serve as theoretical benchmarks, as each constitutes a special case of the general model. Batch auctions correspond to a mechanism $\Gamma$ that disregards individual-trade bids; independent standard auctions, as implemented by 1inch and UniswapX, correspond to a mechanism that disregards batch bids.

\subsection{Independent standard auctions}\label{sec: independent standard}

When both trades are auctioned, solvers internalize the possibility of winning both trades and generating the additional efficiencies from joint execution. In simultaneous auctions with complementarities, this gives rise to what the literature calls the \emph{exposure problem}: a bidder who expects to win one auction with positive probability values winning the other more and may therefore bid more aggressively. By doing so, however, the bidder incurs losses when winning only one auction (see \citealp{meng2017exposure}).

In our setting, the feasibility constraint limits such exposure. Because each order may be canceled, the event in which a solver executes one order alone has strictly positive probability under every strategy profile: solver~1 can never bid more than $\udelta$ for the second order, and solver~2 can never bid more than $\ubeta$ for the first order. Solver~1 therefore bids on order~1 knowing that the opponent's bid on that order is bounded above by $\ubeta$, and solver~2 bids on order~2 knowing that the opponent's bid is bounded above by $\udelta$. The resulting equilibrium is described in the following lemma (its proof follows directly from the preceding argument).

\begin{lemma}\label{lem: dutch auction}
In the independent standard auctions, in every equilibrium of the game solver~1 wins the first order and returns at most $\ubeta$ units of asset~$B$, while solver~2 wins the second order and returns at most $\udelta$ units of asset~$D$.
\end{lemma}

The intuition is immediate in the case of second-price auctions: in each auction, the lowest bid determines the quantity returned to the trader by the winning solver. If the two lowest bids are $\ubeta$ and $\udelta$, these are the amounts traders receive. But because the least productive solver on each trade never wins, there are equilibria in which the two lowest bids are lower than $\ubeta$ and $\udelta$. The same logic extends to other standard auction formats. In particular, the result holds for first-price auctions, which are strategically equivalent to the Dutch auctions used in practice.\footnote{\label{footnote: continuous vs discrete} There are some technicalities arising from the fact that the bid space is continuous. If the bid space was discrete, in a first-price auction the strong solver outbids the weak solver by the minimal increment. Such ``minimal increment'' is not defined when the bid space is continuous. Instead, with a continuous bid space, in the unique equilibrium both solvers bid the same amount, and an endogenous tie-breaking rule (\citealp{jackson2002communication}) assigns the win to the stronger solver. See also \citet{GENTRY2019109}, who use such a rule to establish equilibrium existence in simultaneous auctions with complementarities. Note that, absent cancellation risk, each solver's set of feasible bids would depend on the opponent's strategy; such a game falls outside the framework of \citet{jackson2002communication}, whose endogenous tie-breaking approach requires exogenous action spaces, and the existence results of \citet{GENTRY2019109} would not apply.}

Traders, therefore, receive at most the minimum amount of assets. Intuitively, specialization makes each auction noncompetitive: there is a strong and a weak solver in each, and the strong solver can win by merely matching the weak solver's bid. The productivity of the weak solver in each auction thus determines what traders receive.

\subsection{Batch auction}\label{sec: batch auction}

To start, note that feasibility rules out ``second price'' mechanisms that use the losing bid to determine the assets returned by the winning solver: even if the winning bid has higher total value, it may return 
less of one asset than the second-highest bid. We therefore focus on a ``first-price'' 
(or ``pay-your-bid'') mechanism in which the winning solver delivers the assets specified 
in its batched bid, which is also CoW Protocol's approach.

In a batch auction, each bid is evaluated by the market value of the assets returned to the 
traders. This generates indeterminacy, since the same market value can result from different 
combinations of assets~$B$ and~$D$. We start by considering a specific equilibrium: one in which each solver bids the maximum on its weaker order. In such an equilibrium, $Q^*_2(B)= g \ubeta$ and $Q^*_1(D)=g \udelta$. Define  $V\equiv g \ubeta - p_{DB} \cdot g \udelta$ and write solver 1's best response as
\begin{equation}\label{eq:BR-batch-1}
   \mbox{argmax}_{Q_1(B) \leq g \beta }\left\lbrace \left(\beta \cdot g - Q_1(B) \right) \mbox{pr}\left\lbrace Q_1(B)   >   p_{DB}Q_2(D)  + V  \right\rbrace  \right\rbrace  
\end{equation}
and solver 2's best response as
\begin{equation}\label{eq:BR-batch-2}
\mbox{argmax}_{Q_2(D) \leq g \delta }\left\lbrace p_{DB} \left(\delta \cdot g - Q_2(D) \right) \mbox{pr}\left\lbrace p_{DB}Q_2(D) +V > Q_1(B)      \right\rbrace \right\rbrace .
\end{equation}

We now introduce the following change of variables: $\hat x=Q_1(B)-V$,  and $\hat y=p_{DB} Q_2(D)$. By doing so, we can rewrite the solvers' best responses as
    \[
     \mbox{argmax}_{\hat x} \left\lbrace ( g \beta -V -\hat x) \mbox{pr}\{\hat x \geq \hat y\} \right\rbrace
    \]
    \[
  \mbox{argmax}_{\hat y} \left\lbrace ( p_{DB} \cdot g \delta -{\hat y}) \mbox{pr}\{{\hat y} \geq {\hat x}\} \right\rbrace,
    \]
    which are the two best responses of a first-price auction with types distributed over $[g \ubeta - V, g \overline \beta -V]$ and $[p_{DB} \cdot g \udelta,p_{DB} g \overline \delta]$, leading to the following lemma (its proof follows directly from the preceding argument).

\begin{lemma}\label{lem: batched auction}
There is an equilibrium of the batch auction with bids $Q^*_2(B)= g \ubeta$, $Q^*_1(D)=g \udelta$ and $Q^*_1(B), Q^*_2(D)$, where $\{Q^*_1(B) - V, ~ p_{DB} Q^*_2(D)\}$ are the equilibrium bids of a first price auction with types distributed over $[g \ubeta - V, g \overline \beta -V]$ and $[p_{DB} \cdot g \udelta,p_{DB} \cdot g \overline \delta]$ (appropriately derived from the underlying distribution of the parameters $\beta$ and $\delta$).
\end{lemma}    

As noted above, solvers care only about the total value of their bids, which implies that the 
equilibrium characterized in Lemma~\ref{lem: batched auction} is not unique. Specifically, for 
solver~1, any $\tilde Q_1(B) \in [0, g\beta]$ and $\tilde Q_1(D) \in [0, g\underline{\delta}]$ 
satisfying
\[
\tilde Q_1(B) + p_{DB}\,\tilde Q_1(D) 
  = Q_1^*(B) + p_{DB}\,Q_1^*(D) 
  > g\underline{\beta} + p_{DB}\,g\underline{\delta}
\]
constitutes an equilibrium bid. Similarly, for solver~2, any 
$\tilde Q_2(B) \in [0, g\underline{\beta}]$ and $\tilde Q_2(D) \in [0, g\delta]$ satisfying
\[
\tilde Q_2(B) + p_{DB}\,\tilde Q_2(D) 
  = Q_2^*(B) + p_{DB}\,Q_2^*(D) 
  > g\underline{\beta} + p_{DB}\,g\underline{\delta}
\]
is also an equilibrium bid.

Moreover, the total value returned to traders is at least $g \ubeta + p_{DB} \cdot g \udelta$ in \emph{every} equilibrium of the batch auction, not only in those characterized above. Each solver, when winning both orders, can generate at least $g \ubeta + p_{DB} \cdot g \udelta$ regardless of its type. If the winning batched bid had a lower total value with positive probability, the losing solver could deviate to a feasible batched bid with total value strictly between the winning bid and $g \ubeta + p_{DB} \cdot g \udelta$, winning on those events at a strictly positive margin. No such equilibrium exists.

This bound has a further implication. Suppose solver~1's batched bid wins. Feasibility requires $Q_1(D) \le g \udelta$, so a total value of at least $g \ubeta + p_{DB} \cdot g \udelta$ implies $Q_1(B) \ge g \ubeta$. Symmetrically, if solver~2's batched bid wins, then $Q_2(D) \ge g \udelta$. Hence, in every equilibrium of the batch auction, the trader demanding the asset in which the winning solver is stronger receives at least $g$ times its lower bound.

In sum, in the batch auction, the total value of the assets returned to the traders is at least as high as in the independent individual-trade auctions, and strictly higher whenever $g>1$. The reason is that batching forces the two solvers to compete: there is a single ``strong'' solver per trade and little competition if trades are auctioned off independently, but there are two ``strong'' solvers and high competition in the batch auction. At the same time, in the batch auction, there is an indeterminacy relative to what each trader receives. There are equilibria in which a trader receives less in a batch auction than what this trader would have received in the independent individual-trade auctions. We summarize these observations in the following remark.
\begin{remark}
Evaluated at the notional market price, traders collectively earn at least as much in the batch auction as in the independent standard auctions (cf.\ Lemma \ref{lem: dutch auction}), and strictly more whenever $g>1$. There are equilibria of the batch auction in which both traders are better off relative to the independent standard auctions. However, there are also equilibria in which one of the two traders is worse off --- necessarily, by the argument above, the trader demanding the asset in which the winning solver is weaker.
\end{remark}

The batch auction has equilibria that, intuitively, seem unfair relative to the independent individual-trade auctions. To formalize this intuition, the next section introduces a notion of fairness applicable to combinatorial auctions.

\section{Fair combinatorial auctions}\label{sec: fairness}

The goal of this section is to formalize a notion of fairness applicable to a generic 
combinatorial auction $\Gamma$. We do so by constructing a counterfactual mechanism against 
which to evaluate the outcome of $\Gamma$. 

We begin by defining when an auction's outcome is determined by individual bids.
\begin{defn}[Outcome determined by individual bids]
For a given mechanism $\Gamma$ and message profile
\[
m = \{q_1(B), q_1(D), Q_1(B), Q_1(D), q_2(B), q_2(D), Q_2(B), Q_2(D)\},
\]
we say that the \emph{outcome is determined by the individual bids} if, for all profiles
\[
m' = \{q_1(B), q_1(D), Q'_1(B), Q'_1(D), q_2(B), q_2(D), Q'_2(B), Q'_2(D)\}
\]
such that
\[
Q'_1(B) \le Q_1(B), \quad Q'_1(D) \le Q_1(D), \quad Q'_2(B) \le Q_2(B), \quad Q'_2(D) \le Q_2(D),
\]
it holds that
\[
f(m) = f(m'), \quad 
b_1(m) = b_1(m'), \quad 
d_1(m) = d_1(m'), \quad 
b_2(m) = b_2(m'), \quad 
d_2(m) = d_2(m').
\]
\end{defn}
Intuitively, the outcome is determined by the individual bids when lowering the batched bids componentwise 
does not affect either the matching of trades to solvers or the assets allocated to the traders.

Whenever the outcome is determined by the individual bids, the batched bids can be ignored and we can write
\[
 \tilde b_i(q_1(B), q_1(D), q_2(B), q_2(D)) \equiv b_i(q_1(B), q_1(D), 0, 0, q_2(B), q_2(D), 0, 0) 
          , \quad \forall i \in \{1,2\},
\]
\[
\tilde d_i(q_1(B), q_1(D), q_2(B), q_2(D)) \equiv d_i(q_1(B), q_1(D), 0, 0, q_2(B), q_2(D), 0, 0) 
         , \quad \forall i \in \{1,2\},
\]
where $\tilde b_i, \tilde d_i: \mathbb{R}_+^4 \to \mathbb{R}_+$ also define a 
mechanism. Such a mechanism could correspond, for example, to two independent first-price auctions:
\begin{align}\label{eq: first price}
\begin{split}
\tilde b_i(q_1(B), q_1(D), q_2(B), q_2(D)) &= 
  \begin{cases}
    q_i(B) & \text{if } q_i(B) > q_{-i}(B),\\
    0 & \text{otherwise},
  \end{cases} \\
\tilde d_i(q_1(B), q_1(D), q_2(B), q_2(D)) &=
  \begin{cases}
    q_i(D) & \text{if } q_i(D) > q_{-i}(D),\\
    0 & \text{otherwise},
  \end{cases}
\end{split}
\end{align}
or to two independent second-price auctions:
\begin{align}\label{eq: second price}
\begin{split}
\tilde b_i(q_1(B), q_1(D), q_2(B), q_2(D)) &=
  \begin{cases}
    q_{-i}(B) & \text{if } q_i(B) > q_{-i}(B),\\
    0 & \text{otherwise},
  \end{cases} \\
\tilde d_i(q_1(B), q_1(D), q_2(B), q_2(D)) &=
  \begin{cases}
    q_{-i}(D) & \text{if } q_i(D) > q_{-i}(D),\\
    0 & \text{otherwise}.
  \end{cases}
\end{split}
\end{align}
In other words, whenever the outcome depends solely on individual bids, the mechanism reduces 
to two independent standard auctions that operate as the counterfactual benchmark for fairness. 

We now define a notion of fairness for combinatorial auctions. A mechanism is \emph{fair} if 
its outcome is at least as good for both traders as a counterfactual outcome determined solely 
by the individual bids.

\begin{defn}[Fairness]
A mechanism is \emph{fair} if and only if, for all message profiles 
$m \in \mathbb{R}_+^8$,
\begin{align*}
&\sum_{i=1}^2 b_i(m) \ge 
  \sum_{i=1}^2 \tilde b_i(q_1(B), q_1(D), q_2(B), q_2(D))
\\
&\sum_{i=1}^2 d_i(m) \ge 
  \sum_{i=1}^2 \tilde d_i(q_1(B), q_1(D), q_2(B), q_2(D)).
\end{align*}
\end{defn}
Intuitively, fairness requires that batching never leaves any trader worse off, in terms of the 
assets received, than in the counterfactual mechanism determined only by individual bids.

\medskip
In this definition, fairness is always \emph{relative} to a reference mechanism with payments
$\{\tilde b_i(\cdot), \tilde d_i(\cdot)\}$. The relevance of the notion therefore depends on the choice of 
reference. Any mechanism in which the outcome always depends solely on individual bids is trivially fair. Likewise, the batch auction is 
trivially fair when $\tilde b_i(\cdot) = \tilde d_i(\cdot) \equiv 0$ $\forall i\in\{1,2\}$. A more interesting question is whether 
batching can be fair relative to the outcomes of two non-trivial individual-trade auctions. To answer that question, 
the next sections consider two benchmark cases in which batching arises in equilibrium 
and the fairness reference is constructed from either a first-price auction~\eqref{eq: first price} or a second-price auction~\eqref{eq: second price}.

\medskip
Finally, it is important to emphasize that the reference for fairness is \emph{not} the equilibrium of 
$\{\tilde b_i(\cdot), \tilde d_i(\cdot)\}$. Because the reference depends on individual bids submitted within 
the combinatorial auction, solvers' incentives differ from those in two independent standard 
auctions with payoffs $\{\tilde b_i(\cdot), \tilde d_i(\cdot)\}$, as will be shown below.



\section{Equilibrium of the fair combinatorial auction}\label{sec: equilibrium}

This section derives the equilibrium of two versions of the fair combinatorial auction. 
For ease of comparison with the previous benchmarks, we restrict attention to mechanisms that 
satisfy the following properties:
\begin{itemize}
    \item If the individual bids determine the outcome of the combinatorial auction, the 
    mechanism coincides with two independent first-price auctions, as in 
    \eqref{eq: first price}, or with two independent second-price auctions, as in 
    \eqref{eq: second price}.
    \item  Else the outcome is that of a batched auction. That is, the winning solver pays its bid. 
\end{itemize}

These fair combinatorial auctions introduce several novel strategic considerations. First, solvers may bid more aggressively in the individual auctions to render the opponent's batched bid unfair. 
Conversely, a solver may raise its own batched bid to avoid disqualification on fairness 
grounds. At the same time, if a solver expects the opponent's batched bid to be excluded as 
unfair, it has an incentive to bid less aggressively in the batched auction. Finally, as 
before, the feasibility constraint limits the amount each solver can bid. As shown below, how these  strategic considerations shape the equilibrium is mediated by the auction format, that is, whether the reference for fairness is in first or second price.


\subsection{Second-price individual auctions}\label{sec: secondprice_fair}

When the individual, trade-by-trade auctions follow a second-price 
format, the mechanism is defined as follows:\footnote{As discussed in footnote~\ref{footnote: continuous vs discrete}, when the bidding space 
is continuous, equilibrium existence may depend on tie-breaking. In the fair combinatorial 
auction, ties can occur between a batched bid and an individual-trade bid. When the individual 
auctions are in second price, however, the way ties are broken does not affect the equilibrium. 
This will matter in the next section, when we consider the first-price case.}
\begin{itemize}
    \item If for all $i \in \{1,2\}$, 
    $Q_i(B) \ge \min\{q_1(B), q_2(B)\}$ and 
    $Q_i(D) \ge \min\{q_1(D), q_2(D)\}$, 
    then the outcome is equivalent to a batch auction: 
    $f(m) = \{W, W\}$, $b_W(m) = Q_W(B)$, and $d_W(m) = Q_W(D)$, where 
    \[
W \equiv \mbox{argmax}_{i \in \{1,2\}} \{ Q_i(B) + p_{DB} Q_i(D) \}
\]
is the solver with the highest-valued batched bid, and all other transfers are zero.
    
    \item If there exists a unique solver $i \in \{1,2\}$ such that 
    $Q_i(B) \ge \min\{q_1(B), q_2(B)\}$ and 
    $Q_i(D) \ge \min\{q_1(D), q_2(D)\}$, 
    then the outcome is batching by solver~$i$: 
    $f(m) = \{i, i\}$, $b_i(m) = Q_i(B)$, and $d_i(m) = Q_i(D)$, 
    with all other transfers equal to zero.
    
    \item Otherwise, each order is allocated independently. Let 
    $w(B) \equiv \mbox{argmax}_{i \in \{1,2\}} \{q_i(B)\}$ and 
    $w(D) \equiv \mbox{argmax}_{i \in \{1,2\}} \{q_i(D)\}$ denote the winners of orders~1 and~2, 
    respectively, where the argmax is set-valued in case of ties and is otherwise identified with its unique element. The outcome is then: 
    $f(m) = \{w(B), w(D)\}$, 
    $b_{w(B)}(m) = \min\{q_1(B), q_2(B)\}$, 
    $d_{w(D)}(m) = \min\{q_1(D), q_2(D)\}$, 
    with all other transfers equal to zero.
\end{itemize}

The next proposition shows that the fair combinatorial auction with second-price individual trade auctions provides no fairness guarantees at all.

\begin{proposition}\label{prop: second price}
The set of equilibria of the fair combinatorial auction with second-price individual-trade 
auctions (as in~\eqref{eq: second price}) coincides with the set of equilibria of the batch 
auction.
\end{proposition}

Intuitively, when the individual-trade auctions are in second price, solvers have  limited ability to influence the reference for fairness. For example, if a solver bids zero on the individual trades, then the reference for fairness is zero independently of what the other solver bids. There is therefore an equilibrium in which both solvers bid zero on individual trades and the mechanism reduces to the batch auction 
(cf.\ Lemma~\ref{lem: batched auction}). There are no fairness guarantees.

Finally, note that in this case the mechanism used to define fairness is the second-price auction, yet a trader may receive \textit{less} than in the equilibrium of two independent second-price auctions (Lemma~\ref{lem: dutch auction}). In other words, in equilibrium, the fair combinatorial auction does not guarantee that each trader receives at least as much as they would in the equilibrium of the reference mechanism itself. This lack of guarantee is not a general result: as shown in the next subsection, when the reference for fairness is based on first-price auctions, every trader receives \textit{strictly more} than in the equilibrium of the benchmark mechanism whenever $g>1$.

\subsection{First-price individual auctions}\label{sec:firstprice_fair}

Let $w(B)$ and $w(D)$ denote the solvers with the highest bids on trades~1 and~2, respectively, 
and let $W$ denote the solver with the highest-valued batched bid 
(see the beginning of the previous subsection for definitions). Unlike in the second-price case, 
here the equilibrium depends on how ties are broken. In particular, we must distinguish between 
ties among bids submitted by the same solver and ties among bids from different solvers.

\begin{defn}[Fair batched bid]
The batched bid of solver $i \in \{1,2\}$ is \emph{fair} whenever
\[
\begin{cases}
    Q_i(B) \ge q_{w(B)}(B) & \text{if } i \in w(B),\\
    Q_i(B) > q_{w(B)}(B) & \text{otherwise},
\end{cases}
\qquad\text{and}\qquad
\begin{cases}
    Q_i(D) \ge q_{w(D)}(D) & \text{if } i \in w(D),\\
    Q_i(D) > q_{w(D)}(D) & \text{otherwise.}
\end{cases}
\]
\end{defn}
Hence, a solver's batched bid is \textit{fair} if it matches the individual-trade bids from the same solver, but not if it matches the opponent's individual-trade bids. The mechanism is defined as follows:
\begin{itemize}
    \item If both batched bids are fair, then $f(m) = \{W, W\}$, 
    $b_W(m) = Q_W(B)$, $d_W(m) = Q_W(D)$, and all other transfers are zero.
    \item If only solver $i \in \{1,2\}$ has a fair batched bid, then 
    $f(m) = \{i, i\}$, $b_i(m) = Q_i(B)$, $d_i(m) = Q_i(D)$, and all other transfers are zero.
    \item Otherwise, $f(m) = \{w(B), w(D)\}$, 
    $b_{w(B)}(m) = q_{w(B)}(B)$, $d_{w(D)}(m) = q_{w(D)}(D)$, 
    and all other transfers are zero.
\end{itemize}
The tie-breaking rule is necessary for equilibrium existence. It implies that a solver wishing to make the 
opponent's batched bid \emph{unfair} must match one of its own individual-trade bids to the 
opponent's batched bid. If tie-breaking were different, this solver would need to outbid the 
opponent by an infinitesimal amount, which is undefined in a continuous bid space. Similarly, if 
solver~1's bid on trade~1 is high enough to make solver~2's batched bid unfair, solver~1 may want to 
set $Q_1(B)$ at the minimal level that does not disqualify it as unfair. Under the adopted tie-breaking 
rule, this level is $Q_1(B) = q_1(B)$; under an alternative rule, this minimum is not 
well-defined.

Because each order may be canceled, a solver may have to execute any individually won order alone; Assumption~\ref{ass: feasibility} therefore caps the individual bids at the stand-alone frontiers:
\[
q_1(B) \le \beta, \qquad q_1(D) \le \udelta, \qquad q_2(B) \le \ubeta, \qquad q_2(D) \le \delta,
\]
while batched bids are capped at the joint frontiers. Two consequences follow immediately. First, a solver can hope to disqualify the opponent's batched bid only through the trade in which it is stronger: solver~1's bid on order~2 can never exceed $\udelta$, and hence cannot disqualify any batched bid delivering more than $\udelta$ of asset~$D$ (and symmetrically for solver~2). Second, since the weaker component of solver~2's batched bid is capped at $g\ubeta$, disqualifying it requires $q_1(B) \ge g\ubeta$, which is feasible if and only if $\beta \ge g\ubeta$; symmetrically, solver~2 can disqualify solver~1's batched bid only if $\delta \ge g\udelta$. The ability to filter the opponent's batched bid is therefore a property of a solver's type, with thresholds $g\ubeta$ and $g\udelta$.

The following lemma provides a useful characterization of the equilibrium.
\begin{lemma}\label{lem: intermediate}
If $g>1$, then in every equilibrium of the game it must hold that 
$Q_1^*(D) = g\,\underline{\delta}$ and $Q_1^*(B) \in (\underline{\beta}, g\beta)$, 
while $Q_2^*(B) = g\,\underline{\beta}$ and $Q_2^*(D) \in (\underline{\delta}, g\delta)$. 

If $g = 1$, the equilibrium coincides with that of two independent first-price auctions, 
as described in Lemma~\ref{lem: dutch auction}. Without loss of generality, we may also assume 
$Q_1^*(D) = g\,\underline{\delta}$ and $Q_2^*(B) = g\,\underline{\beta}$ when $g=1$.
\end{lemma}

The lemma implies that, in equilibrium, each solver's batched bid achieves the maximum on this solver weaker trade. When $g>1$, this strategy is unique because there is a strictly 
positive probability that a batched bid wins; when $g=1$, batching never occurs, so any batched bid is an equilibrium bid. Intuitively, a solver's batched bid is more likely to be deemed unfair due to the asset associated with its weaker trade, which is the opponent's 
stronger trade. On the other component of the batched bid, each solver delivers strictly more 
than the minimum feasible amount but less than the maximum.

\begin{footnotesize}
    
 \begin{table}[h!]
\centering

\begin{NiceTabular}{c||c|c|}[hvlines]
&  $q_{1}(B)< g \ubeta $ & $q_{1}(B)\geq  g \ubeta$  \\ 
\hline \hline
  $q_{2}(D) < g \udelta $        & $\begin{cases} (1,1) &\mbox{if } Q_{1}(B) + p_{DB} \cdot g \udelta > g \ubeta + p_{DB}Q_{2}(D) \\ (2,2)  &\mbox{otherwise} \end{cases}$   
&     $(1,1)$          \\ 
$q_{2}(D) \geq g \udelta $                  &   (2,2)            &    $(1,2)$                            \\ 
\end{NiceTabular}

\caption{Matching of trades to solvers as a function of the bids not characterized by Lemma \ref{lem: intermediate}.}
\label{tab:1}
\end{table}

\end{footnotesize}

Combining the bid caps with the above lemma, Table \ref{tab:1} plots the auction outcome as a function of the solvers' bids. We can, therefore, write solver $1$'s payoff as:
\begin{align*}
\begin{cases}
(\beta - q_1(B)) \mbox{pr} \left\lbrace q_2(D) \geq  g \udelta \right\rbrace  +  \left(\beta \cdot g - Q_1(B) \right) \mbox{pr} \left\lbrace q_2(D) < g \udelta \right\rbrace   &\mbox{ if } q_1(B)\geq g \ubeta \\
\left(\beta \cdot g - Q_1(B) \right)  \mbox{pr}\left\lbrace q_2(D) < g \udelta\right\rbrace  \mbox{pr}\left\lbrace Q_1(B)   >   p_{DB}Q_2(D) + V  | q_2(D) < g \udelta  \right\rbrace &\mbox{ if } q_1(B) < g \ubeta
\end{cases}
\end{align*}
where, again, $V\equiv g \ubeta - p_{DB} \cdot g \udelta$. Note that, as long as $q_1(B) \geq  g \ubeta$, whether solver 1 wins one or both trades is independent of its bids, and solver 1 should set $q_1(B)= Q_1(B)=g \ubeta$. A second observation is that if $q_1(B)< g \ubeta$, then solver 1's payoff does not depend on $q_1(B)$ and, without loss of generality, we can assume $q_1(B)=\ubeta$. We can therefore rewrite solver 1's payoff as

\begin{align}\label{eq: payoff solver 1}
\begin{cases}
 (\beta - g \ubeta) \, \mathrm{pr} \{ q_2(D) \geq g \udelta \} + (\beta \cdot g - g \ubeta) \, \mathrm{pr} \{ q_2(D) < g \udelta \} & \text{if } q_1(B) = Q_1(B) = g \ubeta \\
\\
\begin{aligned}
    & (\beta \cdot g - \tilde{Q}_1(B)) \, \mathrm{pr} \{ q_2(D) < g \udelta \} \\
    & \quad \quad \times \, \mathrm{pr} \{ \tilde{Q}_1(B) > p_{DB} Q_2(D) + V \mid q_2(D) < g \udelta \}
\end{aligned}
& \text{if } q_1(B) = \ubeta, Q_1(B) = \tilde{Q}_1(B)
\end{cases}
\end{align}
where 
\[
\tilde Q_1(B) \equiv \mbox{argmax}_{Q_1(B)} \left\lbrace \left(\beta \cdot g - Q_1(B) \right) \mbox{pr}\left\lbrace Q_1(B)   >   p_{DB}Q_2(D)  + V   | q_2(D) < g \udelta   \right\rbrace \right\rbrace .
\]
Following similar steps, by using Lemma \ref{lem: intermediate} and eliminating dominated strategies, we can rewrite solver 2's payoff as
\begin{align}\label{eq: payoff solver 2}
p_{DB} \cdot
\begin{cases}
(\delta - g \udelta) \, \mathrm{pr} \{ q_1(B) \geq g \ubeta \} + (\delta \cdot g - g \udelta) \, \mathrm{pr} \{ q_1(B) < g \ubeta \} & \text{if } q_2(D) = Q_2(D) = g \udelta \\
\\
\begin{aligned}
    & (\delta \cdot g - \tilde{Q}_2(D)) \, \mathrm{pr} \{ q_1(B) < g \ubeta \} \\
    & \quad \quad \times \, \mathrm{pr} \{ p_{DB} Q_2(D) + V > Q_1(B) \mid q_1(B) < g \ubeta \}
\end{aligned}
& \text{if } q_2(D) = \udelta, Q_2(D) = \tilde{Q}_2(D)
\end{cases}
\end{align}
where
\[
\tilde Q_2(D) \equiv \mbox{argmax}_{Q_2(D)} \left\lbrace p_{DB} \left(\delta \cdot g - Q_2(D) \right) \mbox{pr}\left\lbrace p_{DB}Q_2(D) +V > Q_1(B)        | q_1(B) < g \ubeta   \right\rbrace \right\rbrace .
\]
For future reference, note that $\tilde Q_1(B)$ and $\tilde Q_2(D)$ are almost identical to the best responses in a batch auction (see Equations \ref{eq:BR-batch-1} and \ref{eq:BR-batch-2}). The only difference is that, here, the probability distributions are conditional on $q_2(D) < g \udelta $ and $q_1(B) < g \ubeta$.

The following Lemma shows that solver 1 always prefers $q_1(B) =  g \ubeta$ (i.e., filtering out the opponent's batched bid) to  $q_1(B)< g \ubeta$ (i.e., not filtering out the opponent's batched bid). By the bid caps, $q_1(B) = g \ubeta$ is feasible if and only if $\beta \ge g \ubeta$. Similarly, solver 2 sets $q_2(D)=g \udelta$ when feasible, and $q_2(D)= \udelta$ otherwise. 
\begin{lemma}\label{lem: best responses}
In every equilibrium of the game, solver 1 sets:
\[
\begin{cases}
    q_1(B)= Q_1(B)=g \ubeta &\mbox{ if } \beta \geq g \ubeta \\
    q_1(B)=\ubeta, ~ Q_1(B)=\tilde Q_1(B) &\mbox{ otherwise, }
\end{cases}
\]
and solver 2 sets:
\[
\begin{cases}
    q_2(D)= Q_2(D)=g \udelta &\mbox{ if } \delta \geq g \udelta \\
    q_2(D)=\udelta, ~ Q_2(D)=\tilde Q_2(D) &\mbox{ otherwise }
\end{cases}
\]
\end{lemma}

\begin{figure}[ht]
    \centering
    \begin{tikzpicture}[scale=1.2]
    \draw[->] (0,0) -- (10,0) node[anchor=north] {$\beta$};
    \draw[->] (0,0) -- (0,10) node[anchor=east] {$\delta$};
    
    \draw[dashed] (1,0)node[below]{$\ubeta$} -- (1,9);
    \draw[dashed] (5,0)node[below]{$g \cdot \ubeta$} -- (5,9);
    \draw[dashed] (9,0)node[below]{$\overline \beta$} -- (9,9);
    \draw[dashed] (0,1)node[left]{$\udelta$} -- (9,1);
    \draw[dashed] (0,5)node[left]{$g \cdot \udelta$} -- (9,5);
    \draw[dashed] (0,9)node[left]{$\overline \delta$} -- (9,9);
    
    
    \node at (7,8) {\footnotesize $q_1(B)= Q_1(B)=g \ubeta$};
    \node at (7,7) {\footnotesize$q_2(D)= Q_2(D)=g \udelta$};
    \node at (7,6) {\footnotesize (specialization)};

\node at (7,4) {\footnotesize $q_1(B)= Q_1(B)=g \ubeta$};
\node at (7,3) {\footnotesize $q_2(D)= \udelta; ~ Q_2(D)=\tilde Q_2(D)$}; 
\node at (7,2) {\footnotesize (batching solver 1)};

\node at (3,8) {\footnotesize $q_1(B)= \ubeta; ~ Q_1(B)=\tilde Q_1(B)$};
 \node at (3,7) {\footnotesize$q_2(D)= Q_2(D)=g \udelta$};
 \node at (3,6) {\footnotesize (batching solver 2)};

 \node at (3,4) {\footnotesize $q_1(B)= \ubeta; ~ Q_1(B)=\tilde Q_1(B)$};
     \node at (3,3)  {\footnotesize $q_2(D)= \udelta; ~ Q_2(D)=\tilde Q_2(D)$};
\node at (3,2)  {\footnotesize (competitive batching)};
     
\end{tikzpicture}
    \caption{The two best responses (note: we only consider the bids not already pinned down by Lemma \ref{lem: intermediate})}
    \label{fig:best responses}
\end{figure}

Figure~\ref{fig:best responses} illustrates the two best-response functions in the 
$(\beta, \delta)$ space. It identifies four equilibrium regimes --- specialization, competitive batching, and uncompetitive batching, by either solver --- formally described in the following proposition.

\begin{proposition}\label{prop: first price fca}
The equilibrium of the fair combinatorial auction with first-price individual-trade 
auctions (as in~\eqref{eq: first price}) is unique up to payoff-irrelevant components of the bids and the endogenous tie-breaking rule, and is as follows:
\begin{itemize}
    \item \textbf{Competitive Batching:} If $\beta < g\underline{\beta}$ and $\delta < g\underline{\delta}$, then 
    $q_1^*(B) = \underline{\beta}$ and $q_2^*(D) = \underline{\delta}$, and 
    $\{Q_1^*(B) - V,\, p_{DB} Q_2^*(D)\}$ are the equilibrium bids of a first-price auction 
    with types distributed over 
    $[\,g\underline{\beta} - V,\, \min\{g^2\underline{\beta} - V,\, g\overline{\beta} - V\}\,]$ 
    and 
    $[\,p_{DB} g\underline{\delta},\, p_{DB} \min\{g^2\underline{\delta},\, g\overline{\delta}\}\,]$, 
    with distributions derived from the underlying distributions of $\beta$ and $\delta$. 
    The winning solver wins both trades.

    \item \textbf{Specialization:} If $\beta \ge g\underline{\beta}$ and $\delta \ge g\underline{\delta}$, then 
    solver~1 bids $q_1^*(B) = Q_1^*(B) = g\underline{\beta}$ and 
    solver~2 bids $q_2^*(D) = Q_2^*(D) = g\underline{\delta}$. 
    Solver~1 wins trade~1, and solver~2 wins trade~2.

    \item \textbf{Uncompetitive Batching (solver 1):} If $\beta \ge g\underline{\beta}$ and $\delta < g\underline{\delta}$, then 
    solver~1 bids $q_1^*(B) = Q_1^*(B) = g\underline{\beta}$, while 
    solver~2 bids $q_2^*(D) = \underline{\delta}$ and 
    $Q_2^*(D) = \tilde Q_2(D)$. Solver~1 wins both trades.

    \item \textbf{Uncompetitive Batching (solver 2):} If $\beta < g\underline{\beta}$ and $\delta \ge g\underline{\delta}$, then 
    solver~1 bids $q_1^*(B) = \underline{\beta}$ and $Q_1^*(B) = \tilde Q_1(B)$, while 
    solver~2 bids $q_2^*(D) = Q_2^*(D) = g\underline{\delta}$. Solver~2 wins both trades.
\end{itemize}
The remaining equilibrium bids are as characterized in Lemma~\ref{lem: intermediate}.
\end{proposition}

\begin{proof}
The only remaining step is to verify that, in the first case, $\tilde Q_1(B)$ and 
$\tilde Q_2(D)$ are best responses in a first-price auction. It suffices to apply the same 
argument used in Section~\ref{sec: batch auction}: define 
$\hat x = Q_1(B) - V$ and $\hat y = p_{DB} Q_2(D)$, and rewrite the corresponding 
best-response functions.
\end{proof}

In equilibrium, the four cases of the proposition group into three regimes, depending on each solver's ability to use its individual-order bid to exclude the opponent's batched bid on fairness grounds. Solver~1 can exclude solver~2's batched bid whenever its productivity on order~1 is high enough to bid at least $g\underline{\beta}$; similarly, solver~2 can exclude solver~1's batched bid whenever its productivity on order~2 is high enough to bid at least $g\underline{\delta}$. When both inequalities hold, both batched bids are excluded and the outcome is \textit{specialization}. When neither holds, both batched bids remain fair and the outcome is \textit{competitive batching}. When only one inequality holds, the outcome is \textit{uncompetitive batching}: one solver excludes the opponent's batched bid and wins the batch without competition.

With respect to the traders' payoffs, in the \textit{specialization} regime, the equilibrium allocation coincides with that of the independent standard auctions (cf.\ Lemma~\ref{lem: dutch auction}), but traders receive $g\underline{\beta}$ units of asset~$B$ and $g\underline{\delta}$ units of asset~$D$ --- strictly more than in the independent standard auctions whenever $g>1$. These amounts are deliverable even when each solver executes its order alone precisely because, in this regime, $\beta \ge g\underline{\beta}$ and $\delta \ge g\underline{\delta}$. The intuition is that each solver raises its 
individual bid strategically to disqualify the opponent's batched bid as unfair: if either 
solver were to bid below the equilibrium level, the opponent's batched bid would win.

Unlike in the batch auction (see Lemma~\ref{lem: batched auction}), in the 
competitive batching regime of the fair combinatorial auction the equilibrium is 
unique, and traders receive \textit{at least} $g\underline{\beta}$ units of asset~$B$ and 
$g\underline{\delta}$ units of asset~$D$. The intuition is that each solver seeks to prevent its 
batched bid from being disqualified and therefore bids the maximum possible on its weaker trade. 
As a result, each trader receives at least what the less efficient solver at that trade could 
generate when executing both trades.

Finally, the remaining two regimes correspond to \textit{uncompetitive batching}. Here, the 
benefit of specialization lies in an intermediate range, and one solver is substantially more 
productive than the other. This more efficient solver can easily render the opponent's batched 
bid unfair and, anticipating no competition, wins by bidding the minimal amount required for its 
own batched bid to be considered fair.

\begin{table}[t]
\centering
\small
\begin{tabularx}{\textwidth}{@{}p{0.38\textwidth}p{0.27\textwidth}p{0.27\textwidth}@{}}
\toprule
Regime and condition 
& Trader payoffs 
& Aggregate value returned \\
\midrule

Specialization \newline
$\beta \ge g\underline{\beta}$, $\delta \ge g\underline{\delta}$
&
$u=(g\underline{\beta},g\underline{\delta})$
&
$g\underline{\beta}+p_{DB}g\underline{\delta}$ \\

\addlinespace

Competitive batching \newline
$\beta < g\underline{\beta}$, $\delta < g\underline{\delta}$
&
$u=(Q_W(B),Q_W(D))$ \newline
with $u\ge (g\underline{\beta},g\underline{\delta})$
&
$Q_W(B)+p_{DB}Q_W(D)$ \\

\addlinespace

Uncompetitive batching \newline
$\beta \ge g\underline{\beta}$, $\delta < g\underline{\delta}$, or \newline
$\beta < g\underline{\beta}$, $\delta \ge g\underline{\delta}$
&
$u=(g\underline{\beta},g\underline{\delta})$
&
$g\underline{\beta}+p_{DB}g\underline{\delta}$ \\

\bottomrule
\end{tabularx}
\caption{Equilibrium regimes in the fair combinatorial auction with first-price individual auctions.}
\label{tab:first_price_fca_regimes}
\end{table}

Table~\ref{tab:first_price_fca_regimes} summarizes the trader payoffs and aggregate value returned in each regime. 
The key takeaway is that, when the individual auctions are in first price, each solver can directly influence 
the reference for fairness. In equilibrium, both solvers bid more aggressively than in the 
independent individual auctions. The auction therefore provides strong fairness guarantees: 
trader~1 receives at least $g\underline{\beta}$ units of asset~$B$ and trader~2 at least $g\underline{\delta}$ units of asset~$D$, compared with at most $\underline{\beta}$ and $\underline{\delta}$ in the independent first-price auctions 
(cf.\ Lemma~\ref{lem: dutch auction}). In this respect, the mechanism extends to both traders the protection that the batch auction provides only to the trader on the winning solver's stronger side (see Section~\ref{sec: batch auction}). At the same time, the total value returned may fall below the batch auction, namely under specialization and uncompetitive batching. The comparison thus highlights 
a trade-off in the design of fair combinatorial auctions: stronger fairness guarantees come at the expense of lower aggregate value.

The magnitude of this trade-off depends on the parameter $g$, which measures the extra efficiencies from batching. When $g$ is sufficiently large, the equilibrium is almost always competitive batching: the value returned to traders equals that of the batch auction, and the outcome is fair. As $g$ decreases, the total value returned to the 
traders declines while the fairness guarantees are preserved. Intuitively, when the benefit of 
batching is large, those gains can be shared in a way that is fair to both traders; when the 
benefit is small, achieving fairness requires distorting the allocation relative to the simple 
batch auction.

\section{Discussion and extensions}\label{sec: discussions}

This section discusses several features of the model and outlines possible extensions.

\subsection{Benchmark with frictionless financial markets}\label{sec: benchmark}

It is now useful to depart from the main assumption and examine how the problem changes when 
financial markets are perfect, that is, when all agents can frictionlessly exchange assets at 
commonly known market prices.

Let $p_{BA}$ denote the market price of asset~$B$ in terms of asset~$A$, and $p_{DC}$ the price 
of asset~$D$ in terms of asset~$C$. If financial markets are perfect, the equilibrium of the 
independent standard auctions is given by $p_{BA}$ and $p_{DC}$. Each solver can always deliver 
at least $p_{BA}$ and $p_{DC}$ by accessing the financial market, and none would deliver more, 
since any surplus could be sold at those prices. Because $p_{BA}$ and $p_{DC}$ are common 
knowledge, the auction designer also knows the equilibrium of the independent standard auctions. 
To eliminate ``unfair'' equilibria in the batch auction, it suffices to impose limit prices equal 
to $p_{BA}$ and $p_{DC}$.

In summary, the presence of trading frictions that are solvers' private information is a necessary condition for fairness to become a meaningful design problem. When markets are frictionless, the auction designer already knows the benchmark outcomes, whereas under frictions, the reference for fairness must be 
elicited endogenously as part of the mechanism.

\subsection{The fairness--efficiency trade-off}

As already noted, our results illustrate an efficiency--fairness trade-off in the design of fair combinatorial auctions. This trade-off is very different from a standard constrained-efficiency problem. 

If the designer cared only about the aggregate market value returned to traders, the problem would be a standard optimal-auction problem, and the optimal auction would be a batch auction, possibly subject to suitable reserve or limit prices (on optimal auction design, see \cite{klemperer1999auction}). The problem would be similarly straightforward if the fairness benchmark were known exogenously: the optimal auction would solve a constrained optimization problem. 

The distinctive feature of our environment is that the relevant benchmark is not known to the designer. It depends on solvers' private information about routing opportunities, liquidity, and execution costs. The mechanism must therefore cope with a two-dimensional information asymmetry: it must elicit both the reference execution and the execution frontier. Our results highlight a trade-off between these two objectives.

\paragraph{A one-parameter family of fair combinatorial auctions}\label{sec: lambda family}

Our results so far present a very stark view of the fairness--efficiency trade-off in fair combinatorial auctions: either there are no fairness guarantees at all and maximum output is returned to the traders, or there are strong fairness guarantees (i.e., exceeding the equilibrium of the independent standard auctions) at the cost of lower output returned to the traders. It is however possible to make this trade-off less stark by changing how the reference for fairness is constructed. 

Focusing on the fair combinatorial auction with first-price individual-trade auctions (Section~\ref{sec:firstprice_fair}), for $\lambda \in [0,1]$, say that the batched bid of solver $i \in \{1,2\}$ is \emph{$\lambda$-fair} whenever it delivers at least $\lambda$ times the reference quantities constructed there, with the same tie-breaking conventions; the mechanism ignores batched bids that are not $\lambda$-fair. The case $\lambda = 0$ corresponds to the batch auction (or, equivalently, to the fair combinatorial auction with second-price individual-trade auctions); the case $\lambda = 1$ corresponds to the fair combinatorial auction with first-price individual-trade auctions; and $\lambda \in (0,1)$ corresponds to intermediate cases.\footnote{It is theoretically possible to consider $\lambda > 1$, requiring batched bids to strictly improve upon the reference. Doing so, however, does not strengthen the mechanism's guarantees: a solver can then disqualify the opponent's batched bid with individual bids below the $\lambda = 1$ levels, and those bids are what traders receive when batching is filtered out. Moreover, for $\lambda > 1$ a solver's individual bid on its \emph{weaker} order could also disqualify the opponent's batched bid, overturning the composition results of Lemma~\ref{lem: intermediate}. We therefore restrict attention to $\lambda \le 1$.}

\begin{lemma}\label{lem: lambda batch}
If $\lambda \le \min\left\lbrace g\ubeta/\overline\beta,\ g\udelta/\overline\delta \right\rbrace$, the set of equilibrium outcomes of the $\lambda$-fair combinatorial auction coincides with that of the batch auction.
\end{lemma}

A designer who does not know the parameters of the environment but wants to guarantee some level of fairness must therefore choose $\lambda = 1$. For any $\lambda < 1$, there are admissible parameters --- namely, any environment with $g\ubeta/\overline\beta$ and $g\udelta/\overline\delta$ both above $\lambda$ --- for which, by Lemma~\ref{lem: lambda batch}, the mechanism reduces to the batch auction and provides no fairness guarantee at all. At $\lambda = 1$, by contrast, the fairness guarantees of Proposition~\ref{prop: first price fca} hold for all admissible parameters. The fairness--efficiency trade-off documented above is therefore, in part, the price of robustness: mechanisms lower in the family enforce fairness less strictly, but whether they enforce it at all depends on information the designer does not have.

\subsection{Two-stage fair combinatorial auction}

A limitation of the fair combinatorial auction is that solvers submit their batched bids without knowing 
the reference for fairness. Consequently, a solver's batched bid may be disqualified as unfair 
even when a fair batched bid would have been feasible.

A natural solution is to design a \textit{two-stage} fair combinatorial auction. In this 
variant, solvers first submit bids on the individual trades; these bids are then revealed, 
thereby determining the reference outcome for fairness; finally, solvers submit their batched 
bids. However, the two-stage format introduces information leakage: a solver's first-stage 
bids reveal information that its opponent can exploit in the second stage. As a result, in 
equilibrium, solvers submit only \textit{uninformative} bids in the first stage, and the 
mechanism collapses to a standard batch auction, as shown in the following proposition.

\begin{proposition}\label{prop: sequential auction}
In every pure-strategy equilibrium of the two-stage fair combinatorial auction, solver~1 
submits first-stage bids $\hat q_1(B) \le g\underline{\beta}$ and 
$\hat q_1(D) \le g\underline{\delta}$ independent of the realization of~$\beta$. Similarly, 
solver~2 submits fixed first-stage bids $\hat q_2(B) \le g\underline{\beta}$ and 
$\hat q_2(D) \le g\underline{\delta}$ independent of the realization of~$\delta$. In every 
such equilibrium the outcome is batching with probability one, and the winning solver's 
batched bid weakly exceeds the first-stage bids component by component. The set of 
equilibrium outcomes coincides with that of the batch auction.
\end{proposition}

Because the first-stage bids are undetermined, the set of equilibrium outcomes of the two-stage fair combinatorial auction coincides with that of the simple batch auction. The two-stage mechanism therefore admits fair equilibria, but --- like the batch auction itself --- does not guarantee fairness across equilibria.

\subsection{``Unfair'' trade executions}\label{sec: technologically unfair}

A limitation of the baseline model is that unfair outcomes arise solely from the indeterminacy 
in how solvers bid in the batch auction. Consequently, alongside ``unfair'' equilibria, there 
always exist ``fair'' equilibria in which traders are not disadvantaged. This feature is an 
artifact of the model's simplicity. In this section, we show that when solvers have multiple 
ways to execute the two orders, all equilibria of the batch auction may be ``unfair.''

Assume that a solver winning both orders can execute them in two distinct ways. If solver~1 wins 
both orders, it can produce either $\{g\beta,\, g\underline{\delta}\}$ (as in the baseline model) 
or $\{k\underline{\beta},\, \tau\underline{\delta}\}$, where $k> g $ and $\tau < 1$. If 
solver~2 wins both orders, it can produce either $\{g\underline{\beta},\, g\delta\}$ or 
$\{k\underline{\beta},\, \tau\underline{\delta}\}$. We further assume that 
$ g\overline{\beta} < k\underline{\beta}$, so that the alternative production technology 
$\{k\underline{\beta},\, \tau\underline{\delta}\}$ yields more of asset~$B$ but less of 
asset~$D$ relative to both $\{g\beta,\, g\underline{\delta}\}$ and $\{g\underline{\beta},\, g\delta\}$.

Figure~\ref{fig:technologically unfair} illustrates these production technologies for 
solver~1 (right panel) and solver~2 (left panel), together with the total market value of each 
possible execution. The key observation is that, as the relative price $p_{DB}$ varies, the 
technology that maximizes total value when a solver wins both orders also changes. In particular, 
if $p_{DB}$ is sufficiently low, then for both solvers the technology 
$\{k\underline{\beta},\, \tau\underline{\delta}\}$ generates higher value than the alternative. 
The figure also depicts the feasible set of bids under each technology, assuming zero inventory 
for simplicity. Note that when solvers produce $\{k\underline{\beta},\, \tau\underline{\delta}\}$, 
feasibility implies that Trader~2 must receive less than $\underline{\delta}$, resulting in an 
inherently unfair outcome.

\begin{figure}

    \centering

    \hspace*{-1.5cm}    \subfloat[Solver 1's technology.]{
    \begin{tikzpicture}[scale=1.2]

    \draw[->] (0,0) -- (5,0) node[right] {\footnotesize{Asset $B$}};
    \draw[->] (0,0) -- (0,4) node[above] {\footnotesize{Asset $D$}};

    \fill[pattern=vertical lines, pattern color=red] (0,0) rectangle (2.8,.5);
        \fill[pattern=horizontal lines, pattern color=green] (0,0) rectangle (2,2);

    \draw[dotted](0,.5)node[left]{$\tau  \udelta$}--(4.5,.5);
    \draw[dotted](0,1)node[left]{$ \udelta$}--(4.5,1);
    \draw[dotted](0,2)node[left]{$ g  \udelta$}--(4.5,2);

     \draw[dotted](.7,0)node[below]{$\ubeta$}--(.7,3);
    \draw[dotted](1,0)node[below]{$ \beta$}--(1,3);
    \draw[dotted](2,0)node[below]{$ g  \beta$}--(2,3);
     \draw[dotted](2.8,0)node[below]{$ k  \ubeta$}--(2.8,3);

\filldraw (1,1) circle (1pt);
\filldraw[color=red] (2.8,.5) circle (1pt);
\filldraw[color=green] (2,2) circle (1pt);

    \def\slope{-0.75}  

    \draw[dashed] (0,{2 + \slope*(0-2)}) -- ({2 - 2/\slope},0) --({2 - 2/\slope},-.1)node[below] {\footnotesize $g \cdot \beta+ p_{DB} \cdot g \cdot \udelta $};

    \draw[dashed] (0,{.5 + \slope*(0-2.8)}) -- ({2.8 - .5/\slope},0)  -- ({2.8 - .5/\slope},-.5) node[below] {\footnotesize $k \cdot \beta+ p_{DB} \cdot \tau \cdot \udelta $};

\begin{scope}[shift={(0,5)}]

    \draw[->] (0,0) -- (5,0) node[right] {\footnotesize{Asset $B$}};
    \draw[->] (0,0) -- (0,4) node[above] {\footnotesize{Asset $D$}};

    \fill[pattern=vertical lines, pattern color=red] (0,0) rectangle (2.8,.5);
        \fill[pattern=horizontal lines, pattern color=green] (0,0) rectangle (2,2);

    \draw[dotted](0,.5)node[left]{$\tau  \udelta$}--(4.5,.5);
    \draw[dotted](0,1)node[left]{$ \udelta$}--(4.5,1);
    \draw[dotted](0,2)node[left]{$ g  \udelta$}--(4.5,2);

     \draw[dotted](.7,-5)--(.7,3);
    \draw[dotted](1,-5)--(1,3);
    \draw[dotted](2,-5)--(2,3);
     \draw[dotted](2.8,-5)--(2.8,3);

\filldraw (1,1) circle (1pt);
\filldraw[color=red] (2.8,.5) circle (1pt);
\filldraw[color=green] (2,2) circle (1pt);

    \def\slope{-3}  

    \draw[dashed] (1.4,{2 + \slope*(1.4-2)}) -- ({2 - 2/\slope},0) --({2 - 2/\slope},-.1)node[below] {\footnotesize $g \cdot \beta+ p_{DB} \cdot g \cdot \udelta $};

    \draw[dashed] (1.7,{.5 + \slope*(1.7-2.8)}) -- ({2.8 - .5/\slope},0)  -- ({2.8 - .5/\slope},-.5) node[below] {\footnotesize $k \cdot \beta+ p_{DB} \cdot \tau \cdot \udelta $};

\end{scope}

\end{tikzpicture}}
    \subfloat[Solver 2's technology.]{

\begin{tikzpicture}[scale=1.2]

    \draw[->] (0,0) -- (5.5,0) node[right] {\footnotesize{Asset $B$}};
    \draw[->] (0,0) -- (0,4.1) node[above] {\footnotesize{Asset $D$}};

 \fill[pattern=vertical lines, pattern color=red] (0,0) rectangle (2.8,.5);
        \fill[pattern=horizontal lines, pattern color=green] (0,0) rectangle (1.4,3);

    \draw[dotted](0,.5)node[left]{$\tau  \udelta$}--(4.5,.5);
    \draw[dotted](0,1)node[left]{$ \udelta$}--(4.5,1);
    \draw[dotted](0,1.5)node[left]{$ \delta$}--(4.5,1.5);
    \draw[dotted](0,3)node[left]{$ g \cdot \delta$}--(4.5,3);

     \draw[dotted](.7,0)node[below]{$\ubeta$}--(.7,3);
    \draw[dotted](1.4,0)node[below]{$ g  \ubeta$}--(1.4,3);
     \draw[dotted](2.8,0)node[below]{$ k  \ubeta$}--(2.8,3);

\filldraw (.7,1.5) circle (1pt);
\filldraw[color=green] (1.4,3) circle (1pt);
\filldraw[color=red] (2.8,.5) circle (1pt);

    \def\slope{-0.75}  

    \draw[dashed] (0,{3 + \slope*(0-1.4)}) -- ({1.4 - 3/\slope},0) --({1.4 - 3/\slope},-0.1)node[below] {\footnotesize $g \cdot \beta+ p_{DB} \cdot g \cdot \udelta $};

    \draw[dashed] (0,{.5 + \slope*(0-2.8)}) -- ({2.8 - .5/\slope},0)  -- ({2.8 - .5/\slope},-.4) node[below] {\footnotesize $k \cdot \ubeta+ p_{DB} \cdot \tau \cdot \udelta $};

\begin{scope}[shift={(0,5)}]

    \draw[->] (0,0) -- (5.5,0) node[right] {\footnotesize{Asset $B$}};
    \draw[->] (0,0) -- (0,4.1) node[above] {\footnotesize{Asset $D$}};

 \fill[pattern=vertical lines, pattern color=red] (0,0) rectangle (2.8,.5);
        \fill[pattern=horizontal lines, pattern color=green] (0,0) rectangle (1.4,3);

    \draw[dotted](0,.5)node[left]{$\tau  \udelta$}--(4.5,.5);
    \draw[dotted](0,1)node[left]{$ \udelta$}--(4.5,1);
    \draw[dotted](0,1.5)node[left]{$ \delta$}--(4.5,1.5);
    \draw[dotted](0,3)node[left]{$ g  \delta$}--(4.5,3);

     \draw[dotted](.7,-5)--(.7,3);
    \draw[dotted](1.4,-5)--(1.4,3);
     \draw[dotted](2.8,-5)--(2.8,3);

\filldraw (.7,1.5) circle (1pt);
\filldraw[color=green] (1.4,3) circle (1pt);
\filldraw[color=red] (2.8,.5) circle (1pt);

    \def\slope{-2.5}  

    \draw[dashed] (1,{3 + \slope*(1-1.4)}) -- ({1.4 - 3/\slope},0) --({1.4 - 3/\slope},-0.01)node[below] {\footnotesize $g \cdot \beta+ p_{DB} \cdot g \cdot \udelta $};

    \draw[dashed] (1.4,{.5 + \slope*(1.4-2.8)}) -- ({2.8 - .5/\slope},0)  -- ({2.8 - .5/\slope},-.4) node[below] {\footnotesize $k \cdot \ubeta+ p_{DB} \cdot \tau \cdot \udelta $};

\end{scope}

\end{tikzpicture}

    }

    \caption{Solvers' production possibilities when matched with both orders (red and green dots) and when matched with a single order (black dot). The shaded areas are the sets of feasible transfers in case a solver wins both orders, as a function of the choice of production. On the top, $p_{DB}$ is low and $\{k \cdot \ubeta, \tau \udelta\}$ generates higher value than the alternative; on the bottom $p_{DB}$ is high and $\{k \cdot \ubeta, \tau \udelta\}$ generates lower value than the alternative. }
    \label{fig:technologically unfair}
\end{figure}

Under this variation of the model, the equilibrium of the independent standard auctions remains unchanged. The new production possibility arises only when a solver executes both orders jointly, but the feasibility constraint --- via the cancellation risk --- implies that solvers bid on each order as if executing it alone. Consequently, in equilibrium, Trader~1 receives at most $\underline{\beta}$ and Trader~2 at most $\underline{\delta}$ (see Lemma~\ref{lem: dutch auction}). The equilibrium of the batch auction, by contrast, differs markedly. When $p_{DB}$ is 
sufficiently low, both solvers produce $\{k\underline{\beta},\, \tau\underline{\delta}\}$ when 
matched with both orders. Competition then yields $k\underline{\beta}$ for Trader~1 and 
$\tau\underline{\delta}$ for Trader~2. Hence, Trader~2 receives less than $\underline{\delta}$ in 
\textit{all} equilibria, even though it is technologically feasible to produce and allocate 
tokens so that both traders are better off relative to the equilibrium of the independent 
standard auctions. When $p_{DB}$ is low, therefore, \textit{all} equilibria of the batch auction are 
unfair.

Turning to the fair combinatorial auction, when the individual-trade auctions are second price, 
the logic discussed earlier continues to apply: there exists an equilibrium in which both 
solvers set their individual-trade bids to zero, and the mechanism provides no additional 
fairness guarantee relative to the benchmark batch auction. If, instead, the individual-trade 
auctions are first price, the equilibrium again coincides with that derived previously: solvers 
never choose the technology $\{k\underline{\beta},\, \tau\underline{\delta}\}$ because doing so 
would render their batched bids unfair. In this case, both traders receive at least 
$g\underline{\beta}$ and $g\underline{\delta}$, and the mechanism remains effective at providing 
minimal fairness guarantees.

\subsection{Generalization of the fair combinatorial auction to multiple traders: fairness as a filter}

A natural extension concerns how to define the fair combinatorial auction when there are more 
than two traders and more than two solvers. One natural approach is to conduct a combinatorial auction 
in which solvers first submit multiple batched bids (in addition to individual-trade bids), 
then the auctioneer filters out batched bids that are deemed unfair, and finally selects the 
combination of surviving bids that maximizes the total value of tokens returned to traders.

This extension is particularly relevant because fairness affects the computational complexity 
of the mechanism. In standard combinatorial auctions, as the number of trades increases, 
identifying the combination of bids that maximizes aggregate value can become 
computationally intractable---the well-known \textit{winner determination problem}. At the same 
time, if the fairness filter becomes more stringent as the number of trades grows, the 
computational complexity of determining the winners in a fair combinatorial auction may remain 
bounded. A formal analysis of how fairness interacts with computational tractability in 
combinatorial auctions is left for future research.

\section{Conclusion}

We study trade-intent auctions, a new class of trading mechanisms that currently intermediate around USD~9~billion in monthly volume. These auctions emerged in response to the extreme fragmentation of blockchain-based financial markets, which creates three fundamental challenges. First, the best feasible routing an intermediary (in this context, a solver) can provide is private information, as it depends on knowledge of dispersed liquidity across multiple venues and assets, as well as the ability to compute and execute complex routing strategies. This information must therefore be elicited through an auction in which multiple intermediaries compete for the right to execute a trade. Second, joint execution of multiple trades can generate complementarities---for example, through coincidences of wants, sometimes arising in intermediate legs of complex routes---that eliminate execution steps and reduce trading costs. Third, standard best-execution guarantees familiar from traditional finance are absent. While this is partly due to the lack of regulation, it more fundamentally follows from the first challenge discussed above: optimal execution depends on intermediaries' private information, so that best execution cannot be defined independently of the mechanism used to elicit it.

We started by developing a theoretical model of trade-intent auctions, which we used to compare the two 
dominant auction formats: (i) \textit{batch auctions}, in which multiple trade intents are auctioned 
jointly, and (ii) \textit{independent standard auctions}, in which each trade intent is 
auctioned separately (with Dutch auctions as a special case). The main result of this comparison 
is that batch auctions generate stronger competition among solvers than independent standard auctions. Batching forces all solvers to compete: even when each trade has a ``best'' 
solver, the batch auction forces each of them to outbid the other to win any trade. Under the 
assumptions of the model, batch auctions therefore deliver a weakly higher aggregate value to traders than individual trade-intent auctions, strictly so whenever executing multiple trades jointly generates additional efficiencies ($g>1$). At the same time, because solvers' payoffs depend only 
on the total value returned to traders and not on how that value is distributed, the benefits 
of batching may accrue disproportionately to one trader, leaving the other worse off relative 
to the independent standard auctions. This introduces fairness considerations analogous to those explored in cooperative game theory.

Motivated by this observation, we introduce the \textit{fair combinatorial auction}, a 
mechanism that combines elements of batch and independent standard auctions. In the fair combinatorial auction, solvers can submit bids both on individual trades and on batches of trades. The key feature is 
that the outcomes constructed from the individual-trade bids serve as a benchmark for fairness. Batched bids are considered by the auctioneer only if 
they improve upon this benchmark for all traders in the batch; otherwise, they are filtered out 
as unfair. 

The main result shows that, even though the mechanism combines features of batch and 
independent standard auctions, its equilibrium need not combine the equilibria of those 
formats. Solvers anticipate that their individual-trade bids will be used to construct the 
reference for fairness and therefore bid differently in the fair combinatorial auction than in 
independent standard auctions without batching. In designing such an auction, it is crucial to 
account explicitly for these strategic incentives, as they determine the fairness guarantees 
that the mechanism provides \textit{in equilibrium}.

We illustrate this result by showing that when the individual trade-intent auctions are 
second-price auctions, the fair combinatorial auction provides no additional fairness 
guarantees relative to the simple batch auction. By contrast, when the individual trade-intent auctions are first price, each solver can 
influence the reference for fairness through its individual bids, independently of the bids of 
other solvers. In this case, the fair combinatorial auction guarantees that all traders receive strictly more than in the equilibrium of the independent standard auctions. Hence, although not all fair combinatorial auctions provide strong fairness guarantees in equilibrium, some do. These higher fairness guarantees come at the expense of lower value returned to the traders --- a cost that is, in part, the price of robustness: within a one-parameter family of fair combinatorial auctions, the mechanism we propose is the unique member whose guarantees hold in every admissible environment.

\appendix

\section*{Mathematical Appendix}

\begin{proof}[Proof of Proposition \ref{prop: second price}]
    Consider any strictly positive set of bids $\{Q_1(B), Q_2(B), Q_1(D), Q_2(D)\}$ that constitutes an equilibrium of the batch auction discussed earlier. It turns out that these bidding strategies also form an equilibrium of the fair combinatorial auction with second-price individual-trade auctions whenever 
$q_1(B) = q_2(B) \le \min\{Q_1(B), Q_2(B)\}$ and $q_1(D) = q_2(D) \le \min\{Q_1(D), Q_2(D)\}$. The reason is that no individual solver can manipulate the reference for fairness: even if, in equilibrium, $q_1(B) = q_2(B) \le Q_i(B) < \underline{\beta}$ and $q_1(D) = q_2(D) \le Q_i(D) < \underline{\delta}$ for all 
$i \in \{1,2\}$, changing one's individual-trade bids cannot disqualify the opponent's batched bid as unfair. Conversely, no equilibrium delivers the individual-auction outcome: such an outcome requires both batched bids to be ineligible, but each solver can always submit an eligible batched bid --- the thresholds $\min\{q_1(B), q_2(B)\}$ and $\min\{q_1(D), q_2(D)\}$ lie below its feasibility caps --- delivering just above the reference amounts, and whenever $g>1$ such a deviation is strictly profitable. Hence, in every equilibrium the outcome is batching, the fairness constraint is never binding, and the set of equilibria coincides with that of the batch auction, leading to the following lemma.
\end{proof}

\begin{proof}[Proof of Lemma~\ref{lem: intermediate}]
We first show that if $g>1$, then in any equilibrium of the auction 
$Q_1^*(B) > \underline{\beta}$ and $Q_2^*(D) > \underline{\delta}$. We do so by contradiction. 
Suppose that there exists an equilibrium with 
$Q_1^*(B) \le \underline{\beta}$ (the case for solver~2 is analogous). Clearly, this is equivalent to solver~1 not sending a batched bid: solver~2 can disqualify solver~1's batched bid as unfair by bidding $q_2(B) = \ubeta \ge Q_1^*(B)$, and hence there is no equilibrium in which solver~1's batched bid wins. The only possible equilibrium in which solver 1 wins something is one in which $ q_1^*(B) \ge \underline{\beta}$ and solver 1 wins order 1 with strictly positive probability.
 However, if solver~1 deviates and sets 
$Q_1(B) \ge q_1^*(B) \ge \underline{\beta}$ with $Q_1(B) > \underline{\beta}$ and $Q_1(D) = g\,\underline{\delta}$, 
then its batched bid is fair whenever $q_2(D) < g\,\underline{\delta}$ --- an event of strictly positive probability, since $q_2(D) \le \delta$ by feasibility and $\mathrm{pr}\{\delta < g\,\underline{\delta}\} > 0$ when $g>1$ --- and it wins both orders with strictly positive probability. As long as 
$Q_1(B) < g\,\beta$, this deviation is strictly profitable, implying that no equilibrium 
can satisfy $Q_1^*(B) \le \underline{\beta}$.

Next, we show that when $g>1$, it must be that 
$Q_2^*(B) = g\,\underline{\beta}$ and $Q_1^*(D) = g\,\underline{\delta}$. 
Suppose instead that $Q_2^*(B) < g\,\underline{\beta}$. Since 
$Q_2^*(D) > \underline{\delta}$, the probability that solver~2's batched bid is 
considered unfair because of order~2 is zero (solver~1's individual bid on order~2 is 
capped at $\udelta$ by feasibility), while this probability is strictly 
positive for order~1. Solver~2 can therefore improve its payoff by decreasing 
$Q_2^*(D)$ and increasing $Q_2^*(B)$ to keep the total value of the batched bid 
constant, thereby reducing the likelihood of disqualification as unfair. Hence, 
it must be that $Q_2^*(B) = g\,\underline{\beta}$ and $Q_2^*(D) < g\,\delta$. 
By symmetry, $Q_1^*(B) < g\,\beta$ and $Q_1^*(D) = g\,\underline{\delta}$.

Finally, when $g=1$, any feasible batched bid is always considered unfair, so the 
equilibrium coincides with that of two independent first-price auctions. Any 
$Q_2^*(B) \le \underline{\beta}$ and $Q_1^*(D) \le \underline{\delta}$ are equilibrium bids, including 
$Q_2^*(B) = g\,\underline{\beta}$ and $Q_1^*(D) = g\,\underline{\delta}$. 
Thus, whereas for $g>1$ we have 
$Q_2^*(B) = g\,\underline{\beta}$ and $Q_1^*(D) = g\,\underline{\delta}$ in all 
equilibria, for $g=1$ focusing on this case is without loss of generality.
\end{proof}

\begin{proof}[Proof of Lemma \ref{lem: best responses}]
Write \eqref{eq: payoff solver 1} and \eqref{eq: payoff solver 2} as
\begin{align*}
\begin{cases}
\begin{aligned}
X_1  \equiv &(\beta - g \ubeta) \, \mathrm{pr} \{ q_2(D) \geq g \udelta \} \\&+ (\beta \cdot g - g \ubeta) \, \mathrm{pr} \{ q_2(D) < g \udelta \}
\end{aligned}
& \text{if } q_1(B) = Q_1(B) = g \ubeta \\
\\
\begin{aligned}
Y_1  \equiv  & (\beta \cdot g - \tilde{Q}_1(B)) \, \mathrm{pr} \{ q_2(D) < g \udelta \} \\
    & \quad \quad \times \, \mathrm{pr} \{ \tilde{Q}_1(B) > p_{DB} Q_2(D) + V \mid q_2(D) < g \udelta \}
\end{aligned}
& \text{if } q_1(B) = \ubeta, Q_1(B) = \tilde{Q}_1(B)
\end{cases}
\end{align*}

\begin{align*}
\begin{cases}
\begin{aligned}
X_2 \equiv & p_{DB}  \left( (\delta - g \udelta) \, \mathrm{pr} \{ q_1(B) \geq g \ubeta \}\right. \\ & \left. + (\delta \cdot g - g \udelta) \, \mathrm{pr} \{ q_1(B) < g \ubeta \} \right) 
\end{aligned}
& \text{if } q_2(D) = Q_2(D) = g \udelta \\
\\
 \begin{aligned}
 Y_2 \equiv    & p_{DB} \left( (\delta \cdot g - \tilde{Q}_2(D)) \, \mathrm{pr} \{ q_1(B) < g \ubeta \} \right. \\
    & \left. \quad  \times \, \mathrm{pr} \{ p_{DB} Q_2(D) + V > Q_1(B) \mid q_1(B) < g \ubeta \} \right)
\end{aligned}
& \text{if } q_2(D) = \udelta, Q_2(D) = \tilde{Q}_2(D)
\end{cases}
\end{align*}
We want to show that in every equilibrium of the game $X_1 > Y_1$ and $X_2 > Y_2$: in every equilibrium of the game each solver prefers to disqualify the opponent's batched bid as unfair.

We proceed by contradiction. Suppose first that  $X_1 \leq  Y_1$ and $X_2 \leq Y_2$. In this case, the equilibrium is identical to a standard batch auction, but for the fact that $Q^*_1(D)$ and $Q^*_2(B)$ are pinned down by Lemma \ref{lem: intermediate}. By the argument in Section~\ref{sec: batch auction}, in every equilibrium of the batch auction the winning batched bid has total value at least $g \cdot \ubeta + p_{DB} \cdot g \cdot \udelta$. The same bound applies to each solver's bid: a batched bid of lower total value never wins, and raising it to just above $g \cdot \ubeta + p_{DB} \cdot g \cdot \udelta$ --- feasible for every type --- wins with strictly positive probability at a strictly positive margin. Hence, for every realization of $\beta$ and $\delta$, 
\[Q^*_1(B)+p_{DB} Q^*_1(D) \geq g \cdot \ubeta + p_{DB} \cdot g \cdot \udelta\]
and  
\[Q^*_2(B)+p_{DB} Q^*_2(D) \geq g \cdot \ubeta + p_{DB} \cdot g \cdot \udelta.\]
But together with Lemma \ref{lem: intermediate}, the above conditions imply  $Q^*_1(B)\geq g \cdot \ubeta $ and  $ Q^*_2(D) \geq g \cdot \udelta$, which then imply  $X_1 > Y_1$ and $X_2 > Y_2$, therefore contradicting our initial hypothesis. 

Suppose now that $X_1 \leq  Y_1$ and $X_2 > Y_2$ (the argument is identical for the case $X_1 >  Y_1$ and $X_2 \leq Y_2$). If $\delta \geq g \udelta  $, then solver~2 sets $q_2(D)=g\udelta$ and wins both trades. If $\delta < g \udelta  $, then the equilibrium is identical to a batch auction, but for the fact that solver 2's type distribution is truncated from above. Nonetheless, the same argument applies: in every equilibrium, each solver's batched bid has total value at least $g \cdot \ubeta + p_{DB} \cdot g \cdot \udelta$. Again, this implies that $Q^*_1(B)\geq g \cdot \ubeta $ and  $ Q^*_2(D) \geq g \cdot \udelta$, which then imply  $X_1 > Y_1$ and $X_2 > Y_2$, therefore contradicting our initial hypothesis.
\end{proof}

\begin{proof}[Proof of Lemma \ref{lem: lambda batch}]
Recall that the individual bids are capped at the stand-alone frontiers: $q_1(B) \le \beta$, $q_2(B) \le \ubeta$, $q_1(D) \le \udelta$, and $q_2(D) \le \delta$. Hence every individual bid on order~1 is at most $\overline\beta$ and every individual bid on order~2 is at most $\overline\delta$, for every realization of the types. Since the reference quantities are constructed from the individual bids, the $\lambda$-scaled reference is at most $\lambda \overline\beta \le g\ubeta$ on order~1 and at most $\lambda \overline\delta \le g\udelta$ on order~2, where the last inequalities follow from the hypothesis of the lemma. These are precisely the feasibility caps on the weaker components of the batched bids: solver~2 can always bid $g\ubeta$ on order~1 and solver~1 can always bid $g\udelta$ on order~2, while the stronger components exceed these levels for every type. Each solver can therefore always render its batched bid $\lambda$-fair, regardless of the opponent's individual bids. The argument of Section~\ref{sec: secondprice_fair} then applies: no solver can manipulate the reference enough to disqualify the opponent's batched bid, and the set of equilibrium outcomes coincides with that of the batch auction.
\end{proof}

\begin{proof}[Proof of Proposition \ref{prop: sequential auction}]
   First, we prove that, in any pure-strategy equilibrium, solvers bid fixed amounts in the first bidding stage. We do so by contradiction. Suppose that, in equilibrium, solver~1's first-stage bid is not constant in $\beta$. Upon observing solver~1's first-period bid, solver~2 can then update its beliefs about $\beta$, infer the maximum batched bid that solver~1 can submit, and try to outbid it by a minimal amount: solver~2's batched bid is more aggressive the higher the inferred $\beta$. Anticipating this, every type of solver~1 strictly gains by submitting the first-stage bid associated with the lowest inferred $\beta$, thereby deceiving solver~2 into submitting a low batched bid. Hence, all types of solver~1 submit the same first-stage bid, contradicting the assumption that the bid is not constant. An identical argument rules out equilibria in which solver~2's first-stage bid varies with $\delta$.\footnote{We restrict attention to pure-strategy equilibria. We conjecture that the argument extends to mixed strategies---a mixed first-stage bid does not reveal a solver's type exactly, but still lets the opponent update its beliefs and adjust its batched bid accordingly---but we do not prove this here.}

To close the proposition, note that if all first-stage bids are strictly below $g \ubeta$ and $g \udelta$, then with probability 1 there is batching in equilibrium. The first-stage bids do not matter, and feasibility is never violated. If a first-stage bid is strictly greater than $g \ubeta$ or $g \udelta$, there are values of $\beta$ and $\delta$ for which the feasibility constraint is violated. Such bids would, therefore, be informative of the realization of $\beta$ and $\delta$, which cannot be an equilibrium. It remains to consider first-stage bids equal to the bounds. Suppose $\hat q_1(B) = g\ubeta$ (the other cases are analogous). Fairness then requires $Q_2(B) \ge g\ubeta$, which coincides with solver~2's feasibility bound, so $Q_2(B) = g\ubeta$. Such a bid is feasible for every realization of $\delta$: batching occurs with probability one, the first-stage bids are never executed, and feasibility is never violated. The second stage is a batch auction in which the composition of each batched bid is constrained; because the batch auction's winner is determined by the total value of the bids alone, any equilibrium of this constrained game is also an equilibrium of the unconstrained batch auction.
\end{proof}

\bibliography{bib}{}
\bibliographystyle{chicago}

\end{document}